\documentclass[12pt,preprint]{aastex}

\begin{document}


\title{The Oxygen Abundance of Nearby Galaxies from Sloan Digital Sky Survey spectra}

\author{Leonid S. Pilyugin}
\affil{ Main Astronomical Observatory
                 of National Academy of Sciences of Ukraine,
                 27 Zabolotnogo str., 03680 Kiev, Ukraine}
\email{pilyugin@mao.kiev.ua}
\and
\author{Trinh X. Thuan}
\affil{Astronomy Department, University of Virginia, 
P.O. Box 400325, Charlottesville, VA 22904-4325} 
\email{txt@virginia.edu}

\begin{abstract}
We have derived the oxygen abundance for 
a sample of nearby galaxies in 
the Data Release 5 of the Sloan Digital Sky Survey (SDSS)
which possess at least two independent spectra of one or several 
H\,{\sc ii} regions with a 
detected [OIII]$\lambda$4363 auroral line.  
Since, for nearby galaxies, the [OII]$\lambda$3727 nebular line is out of the 
observed wavelength range,  
we propose a method to derive (O/H)$_{\rm ff}$ abundances
using the classic T$_{\rm e}$ method coupled with the ff relation. 
(O/H)$_{7325}$ abundances have also been determined, based on the 
[OII]$\lambda$$\lambda$7320,7330 line intensities, and using  
a small modification of the standard T$_{\rm e}$ method.
The (O/H)$_{\rm ff}$ and  (O/H)$_{7325}$ abundances 
have been derived with both the one- and two-dimensional 
t$_2$ -- t$_3$ relations. 
It was found that the (O/H)$_{\rm ff}$ abundances derived with the parametric 
two-dimensional t$_2$ -- t$_3$ relation are most reliable. 
Oxygen abundances have been determined in 29 nearby galaxies, 
based on 84 individual abundance determinations in H\,{\sc ii} regions. 
Because of our selection methods, the metallicity of our galaxies lies in 
the narrow range 8.2 $\la$ 12 + log (O/H) $\la$ 8.4.  
The radial distribution of oxygen abundances in the disk of the  
spiral galaxy NGC 4490 is determined for the first time. 

\end{abstract}

\keywords{galaxies: abundances -- ISM: abundances -- H\,{\sc ii} regions}

\section{INTRODUCTION}

The chemical composition of a galaxy is one of its most fundamental 
characteristics. Because a galaxy's chemical enrichment 
depends on various physical 
processes, such as the star formation history, the mass exchange between 
a galaxy and its environments and the stellar initial mass function, 
progress in our understanding of galaxy formation and evolution processes 
depends in large part upon improving   
our knowledge of the chemical properties of galaxies. 

Relatively detailed studies of the chemical 
composition of the interstellar medium of galaxies 
can be carried out by using the emission 
lines in spectra of individual H\,{\sc ii} regions in nearby galaxies.
However, such data is very limited. For example,  
spectra of individual 
H\,{\sc ii} regions in spiral galaxies have been obtained  
for only about five dozens objects (see the 
compilation in \citet{pvc04,plato07}). 
The Sloan Digital Sky Survey \citep{yorketal00} provides a very large data 
base of galaxy spectra obtained in a homogeneous way, which allows 
to increase the number of abundance determinations in 
individual H\,{\sc ii} regions in nearby galaxies.
The flux of the auroral oxygen line [OIII]$\lambda$4363 is measurable 
in the SDSS spectra of many H\,{\sc ii} regions 
\citep{izotovetal04,izotovetal06,kniazevetal04}. In those spectra, the
electron temperature t$_3$ of the O$^{++}$ zone can be measured directly.
However, there is a problem concerning 
abundance determination of H\,{\sc ii} regions in nearby galaxies 
with SDSS spectra. 
The observed wavelength range is 3800 -- 9300 \AA\ so that
for nearby galaxies 
with redshifts z $\la$ 0.024, the [OII]$\lambda$3727
emission line is out of that range. 
The absence of the [OII]$\lambda$3727 line prevents the use of the standard 
T$_{e}$ method and of 
empirically or theoretically calibrated relations between 
metallicity and the relative fluxes of strong 
oxygen lines \citep[][among others]{pageletal79,edmundspagel84,dopitaevans86,
mrs85,mcgaugh91,lcal,hcal,calrev} 
to determine oxygen abundances.

It has been suggested by \citet{kniazevetal03,kniazevetal04} 
that this problem can be solved by 
using a modification of the standard T$_{e}$ method, based on the 
intensities of the [OII]$\lambda$$\lambda$7320,7330 
auroral lines, 
instead of the intensity of the [OII]$\lambda$3727 nebular line. 
In this modified version, 
the flux in the [OII]$\lambda$3727 emission line is
estimated from the fluxes of the [OII]$\lambda$$\lambda$7320,7330  
emission lines and the t$_2$ -- t$_3$ relation, where 
 t$_2$ and t$_3$ are respectively 
the electron temperatures of the O$^{+}$ and O$^{++}$ zones.  
However, the [OII]$\lambda$$\lambda$7320,7330 
lines are weak and their measured fluxes usually have large errors, so that  
 the O$^+$ abundances derived from those lines are generally 
not very accurate \citep{izotovetal06}. 
Furthermore, the difference between the t$_2$ temperature derived from the 
[OII]$\lambda$3727/[OII]$\lambda$$\lambda$7320,7330 ratio and the one 
estimated from the commonly used t$_2$ -- t$_3$ relation   
can be very large
\citep[see the discussion in][and references therein]{kennicuttetal03},
making the oxygen abundances so derived even more uncertain.  

Recently, \citet{ff,ohmax} have 
derived a relation, called the ff relation,
 between the auroral [OIII]$\lambda$4363 and nebular
[OIII]$\lambda$$\lambda$4959,5007, [OII]$\lambda$3727 oxygen line fluxes. 
Those authors showed that, by coupling the ff relation with 
 the T$_e$ method, accurate oxygen abundances can be derived 
from measurements of just two oxygen lines,
[OIII]$\lambda$$\lambda$4959,5007 and [OII]$\lambda$3727. 
Can the same combination of the ff relation and the T$_e$ method yield also  
reliable oxygen abundances when only measurements of the two
[OIII]$\lambda$$\lambda$4959,5007 and [OIII]$\lambda$4363 oxygen lines 
are available? We will examine this question here. 
We will also look at the influence of the choice of a particular  
t$_2$ -- t$_3$ relation on the derived abundances.  
After assessing the accuracy of the above method, we will use it   
to determine reliable oxygen abundances in a sample of nearby galaxies
using SDSS spectra.

We describe the spectroscopic data set which forms the basis of the present
investigation in Section 2. 
In Section 3, we discuss the methods used to determine oxygen abundances in  
galactic H\,{\sc ii} regions. 
In Section 4, we examine the reliability of the oxygen abundances determined 
in different ways, using the radial distribution of the oxygen abundance in 
the disk of the well-studied spiral galaxy M101 as a ``Rosetta stone''. 
The oxygen abundances derived in a sample of 
nearby galaxies are discussed in Section 5. 
We summarize our conclusions in Section 6.

For the line fluxes, we will be using the following notations throughout the paper: 
R$_2$ = $I_{[OII] \lambda 3727+ \lambda 3729} /I_{H\beta }$,
R$_3$ = $I_{[OIII] \lambda 4959+ \lambda 5007} /I_{H\beta }$, 
R = $I_{[OIII] \lambda 4363} /I_{H\beta }$, 
R$_{23}$ = R$_2$ + R$_3$. With these definitions,  
the excitation parameter P can then be expressed as: P = R$_3$/(R$_2$+R$_3$).

\section{THE DATA}

\subsection{The galaxy sample}

We have extracted from the Data Release 5 (DR5) 
of the Sloan Digital Sky Survey (SDSS) spectra of    
nearby galaxies which satisfy the following three conditions: 1) For each 
galaxy, there exists at least two independent SDSS spectral measurements 
of one or several H\,{\sc ii} regions in it. This criterion allows us 
to compare the oxygen abundances determined independently  
 and assess their reliability. If there are  
H\,{\sc ii} regions at different galactocentric distances, an abundance 
gradient can also be determined; 2) each H\,{\sc ii} region has 
a high metallicity, i.e. it has an 
oxygen abundance 12 + log (O/H) $\ga$ 8.2; and  
 3) each spectrum shows a [OIII]$\lambda$4363 auroral line. 
The last two criteria select out galaxies that are suited to our 
adopted method of abundance determination, described below.

These three criteria select out about a hundred galaxies out of the 
original 750 000 spectra in the DR5. However, the final list,
given in  Table \ref{table:sample},
includes only those galaxies which we judge to have 
reliable oxygen abundance determinations (our reliability tests are 
discussed in section 4.2), a total of 29 nearby galaxies.
Table \ref{table:sample} gives the general characteristics of each 
galaxy. The first column gives its name. We have used the 
most widely used name for each galaxy. 
The other informations concerning each galaxy 
are taken from the Lyon/Meudon Extragalactic Database (LEDA) and/or 
from the NASA/IPAC Extragalactic Database (NED).
The right ascension and declination (J2000.0) of each galaxy 
are given in column 2, 
its morphological type and morphological type code in columns 3 and 4, 
its apparent B magnitude in column 5,  
its redshift in column 6, and its 
absolute blue magnitude in column 7. A Hubble constant 
of 73 km s$^{-1}$ Mpc$^{-1}$ has been adopted to convert  
from redshift to distance.
Other designations of the galaxy are given in column 8.

Examination of Table \ref{table:sample} shows that our sample 
includes mainly late-type spirals and irregular 
galaxies. Several galaxies do not possess a published 
morphological type.
The majority of the galaxies in the 
sample have a redshift less than 0.01, 
i.e. a distance less than $\sim$ 45 Mpc.
The most distant galaxy in our list, HS1103+4346, has a redshift of 0.021
that corresponds to a distance of $\sim$ 90 Mpc. Their B absolute magnitudes 
range between -21.5, typical of spiral galaxies, and -16.5, typical of 
dwarf irregular galaxies.

\subsection{The line intensity data}

The line intensities of the H\,{\sc ii} regions  
in our sample of nearby galaxies have been derived using 
SDSS spectra in the following way, as illustrated by the following example 
concerning the H${\alpha}$ line. 

The continuum flux level in the 
wavelength range from $\lambda_a$ = 6500 \AA\ to  $\lambda_b$ = 6650 \AA\ 
is approximated by the linear expression 
\begin{eqnarray}
f_{c}(\lambda)= c_0 + c_1 \lambda.
\label{equation:contin}
\end{eqnarray}
The values of the coefficients in Eq.(\ref{equation:contin}) are derived by 
an iteration procedure. In the first step, we exclude the H${\alpha}$ 
line region,  
from 6540 \AA\ to 6590 \AA,  and use all other data points 
to derive a first set of coefficients, using 
the least-squares method. Then, the point 
with the largest deviation is rejected, and a new set of coefficients 
is derived. 
The iteration procedure is continued 
until the differences between two successive values of 
f$_c$($\lambda_a$) and f$_c$($\lambda_b$) are respectively 
less than 0.01f$_c$($\lambda_a$) and 0.01f$_c$($\lambda_b$).

\begin{figure}
\begin{center}
\resizebox{1.00\hsize}{!}{\includegraphics[angle=0]{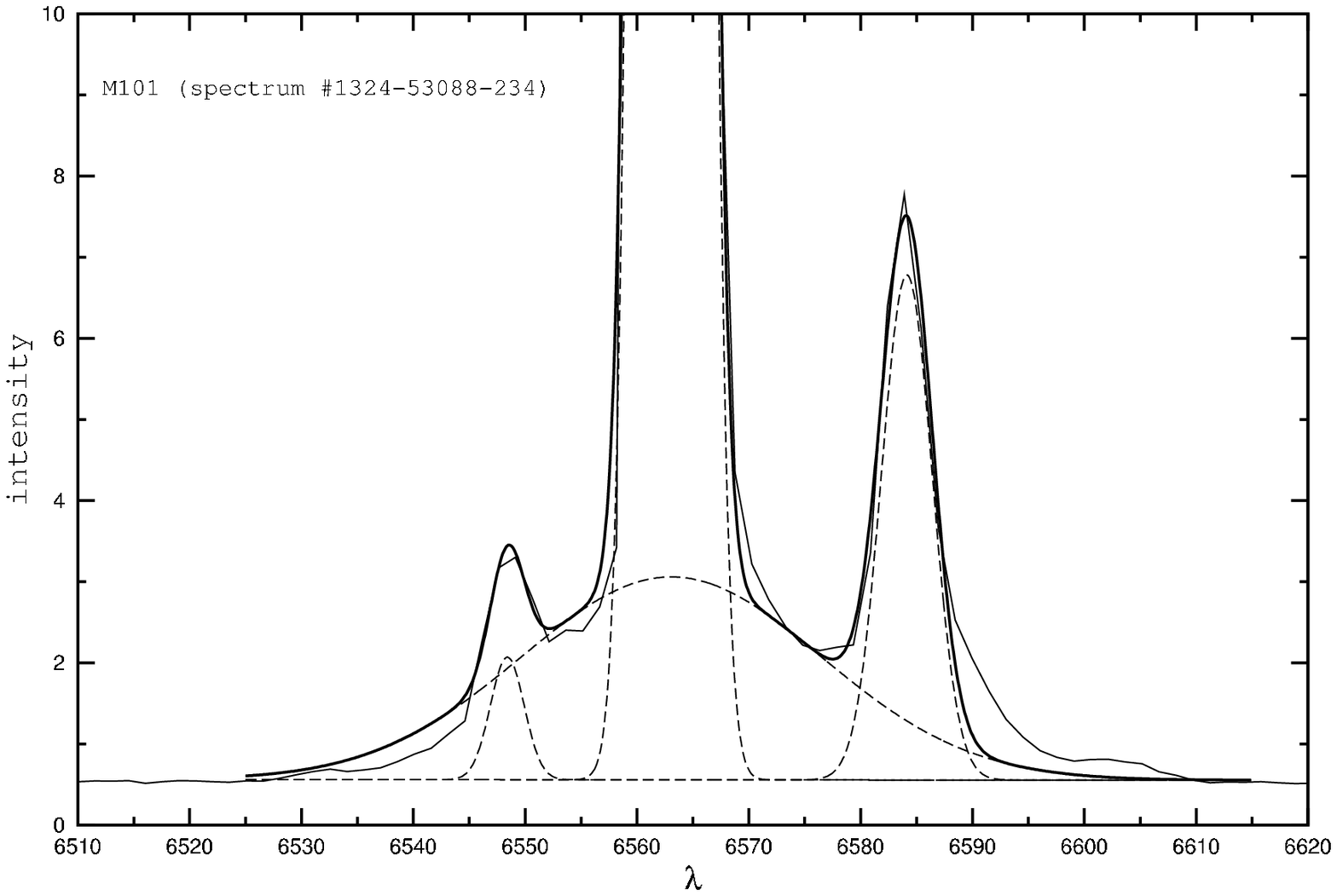}}
\caption{ 
Example of a fit to the [NII]$\lambda$6548, H$\alpha$ (both narrow and
broad components) and  [NII]$\lambda$6584 line profiles.
We show here the spectrum \# 1324-53088-234 of an  
H\,{\sc ii} region in the disk of the spiral galaxy M101. 
The thin solid line is the observed spectrum, the thick solid line is the fit
to it,
and the dashed lines are fits to individual lines. 
}
\label{figure:fit}
\end{center}
\end{figure}

The profile of each line is approximated by a Gaussian of the form
\begin{eqnarray}
f(\lambda)= F\, \frac{1}{\sqrt{2\pi}\,\sigma} 
e^{-(\lambda - \lambda_0)^2/2\sigma^2}
\label{equation:gauss}
\end{eqnarray}
where
$\lambda_0$ is the line central wavelength, 
$\sigma$ is the width of the line, and 
F is the flux in the emission line.
If there is a broad emission, such as in the spectrum \# 1324-53088-234 
shown in Fig.~\ref{figure:fit}
of an H\,{\sc ii} region in the disk of the spiral galaxy M101,   
then several lines are fitted simultaneously. 
In this case,  
the total flux at a fixed value of $\lambda$ is given by the expression
\begin{equation}
f(\lambda)= f_{H\alpha , narrow}(\lambda) + f_{H\alpha , broad}(\lambda)+
f_{[NII]6548}(\lambda) + f_{[NII]6584}(\lambda) + f_{c}(\lambda)
\label{equation:ftot}
\end{equation}
The values of 
F$(H_{\alpha},narrow)$, $\lambda_0(H_{\alpha},narrow)$, $\sigma(H_{\alpha},narrow)$, 
F$(H_{\alpha},broad)$, $\lambda_0(H_{\alpha},broad)$, $\sigma(H_{\alpha},broad)$, 
F([NII]$\lambda$6548, $\lambda_0$([NII]$\lambda$6548), $\sigma$([NII]$\lambda$6548),
F([NII]$\lambda$6584, $\lambda_0$([NII]$\lambda$6584), $\sigma$([NII]$\lambda$65484 
 are derived by requiring that the mean difference 
\begin{eqnarray}
\epsilon = \sqrt{ \frac{1}{n} \sum_{k=1}^{k=n} (f(\lambda_k)-f^{obs}(\lambda_k))^2}
\label{equation:df}
\end{eqnarray}
between the measured flux f$^{obs}(\lambda_k)$ and 
the flux f($\lambda_k$) given by Eq.(\ref{equation:ftot}) 
to be minimum in the range $\lambda_a$ --  $\lambda_b$.
The derived fit to the H${\alpha}$, [NII]$\lambda$6548 and 
[NII]$\lambda$6584 lines is shown by the solid line in Fig.~\ref{figure:fit}. 
A similar procedure is adopted in the 
cases of broad absorption components 
or of partly overlapping lines.  

The measured emission fluxes F are then corrected for interstellar reddening 
using the theoretical H$\alpha$ to H$\beta$ ratio and the analytical 
approximation to the Whitford  
interstellar reddening law from \citet{izotovetal94}. 
In a few cases, the derived 
value of the extinction C(H$\beta$) is negative and is set to zero.

The value of $\epsilon$ (Eq. 4) 
can be considered as an estimate of the uncertainty in the measurement 
of the flux in a 1 {\AA } spectral interval. Since the total 
width of line is approximately equal to 4$\sigma$, the uncertainty in the 
line intensity measurement can be estimated as 
dF = 4$\sigma$$\epsilon$.

\section{ABUNDANCE DETERMINATION}

\subsection{The standard (O/H)$_{\rm st}$ abundances}

To convert line intensities into abundances, 
we adopt a two-zone model for the temperature structure within 
the H\,{\sc ii} region. 
We follow the procedures of \citet{izotovetal06} who 
have recently published a set 
of equations for the determination of oxygen abundances 
in H\,{\sc ii} regions, based on a five-level atom model. 
According to those authors,   
the electron temperature $t_3$ within the [O\,{\sc iii}] zone, in units of 10$^4$K, 
is given by the following equation 
\begin{eqnarray}
t_3 = \frac{1.432}{\log (R_{3}/R)  - \log C_{\rm T}}
\label{equation:t3}
\end{eqnarray}
where
\begin{equation}
C_{\rm T} = (8.44 - 1.09\,t_3 + 0.5\,t_3^2 - 0.08\,t_3^3) \, v 
\end{equation}
\begin{equation}
v = \frac{1 + 0.0004\,x_3}{1 + 0.044\,x_3}  
\end{equation}
and
\begin{equation}
x_3= 10^{-4} n_{\rm e} t_3^{-1/2}.
\end{equation}

As for the ionic oxygen abundances, they are derived from the following equations 
\begin{eqnarray}
12+ \log (O^{++}/H^+) = \log (R_3)  +
  \nonumber  \\
6.200 + \frac{1.251}{t_3}  - 0.55 \log t_3 - 0.014\,t_3,
\label{equation:o3}
\end{eqnarray}

\begin{eqnarray}
12+ \log (O^{+}/H^+) & = & \log (R_2) + 5.961+ \frac{1.676}{t_2} \nonumber  \\
                     & - & 0.40 \log t_2  -0.034\,t_2 \nonumber  \\
                     & + & \log (1+1.35x_2)  ,
\label{equation:o2}
\end{eqnarray}
or
\begin{eqnarray}
12+ \log (O^{+}/H^+) & = & \log (I_{[OII] \lambda 7320+ \lambda 7330}/I_{ H_{\beta} }) \nonumber  \\
                     & + & 6.901+ \frac{2.487}{t_2}  - 0.483 \log t_2   \nonumber  \\
                     & - & 0.013\,t_2 + \log (1-3.48x_2)  ,
\label{equation:o2i}
\end{eqnarray}
where
\begin{equation}
x_2= 10^{-4} n_{\rm e} t_2^{-1/2}.
\end{equation}
Here $n_e$ is the electron density in cm$^{-3}$. The 
determination of the electron temperature t$_2$ will be discussed in 
detail in Section 4.

The total oxygen abundances are then derived from the following equation 
\begin{equation}
\frac{O}{H} = \frac{O^+}{H^+} + \frac{O^{++}}{H^+}                .
\label{equation:otot}
\end{equation}

The oxygen abundances so derived 
from the measured R$_3$ and R$_2$ line intensity ratios, 
using Eqs.(\ref{equation:o3}) and (\ref{equation:o2}), will be referred to 
hereafter as (O/H)$_{\rm st}$ standard abundances. 

\subsection{The (O/H)$_{7325}$ abundances}

As mentioned before, 
because the SDSS spectra cover the 3800 -- 9300A spectral range, the
[OII]$\lambda$3727 emission line is out of the observed 
range for galaxies 
with redshifts z $\la$ 0.024. 
In those cases, O$^+$/H$^+$ cannot be determined by the standard method, 
using Eq.(\ref{equation:o2}). 
To get around that problem, 
\citet{kniazevetal03,kniazevetal04} have suggested a slight modification 
of the standard method. From very general
considerations, it is expected \citep{aller84} that the O$^+$/H$^+$ abundance 
can equally well be derived from the intensities of the 
[OII]$\lambda$$\lambda$7320,7330 auroral lines, 
using Eqs.(\ref{equation:o3}) and (\ref{equation:o2i}).
\citet{kniazevetal04} found that this modified method 
works well over the range of abundances studied by them.
We will refer to abundances derived by that method  
as (O/H)$_{7325}$ abundances. 

\citet{kennicuttetal03} have used respectively 
the [OIII]$\lambda$4363/[OIII]$\lambda$$\lambda$4959,5007 and 
[OII]$\lambda$$\lambda$7320,7330/[OII]$\lambda$3727 auroral to 
nebular line intensity ratios to derive  
the electron temperatures t$_3$ and t$_2$ for a number of relatively 
high-metallicity H\,{\sc ii} regions. 
Comparing t$_2$ to t$_3$, they found the surprising result that 
the two temperatures are uncorrelated for most of the objects in 
their sample. 
They noted that, while the cause of the absence of correlation 
is not known, a possible reason is  
the contribution of recombination processes to the population 
of the level
giving rise to the [OII]$\lambda$$\lambda$7320,7330 lines. 
On the other hand, \citet{izotovetal06} found a correlation between the 
t$_2$ and 
t$_3$ they derived from the above ratios for a sample of low-metallicity  
H\,{\sc ii} regions. Their t$_2$ -- t$_3$ relation follows the one  
predicted by photoionization models, but the scatter of the data points is 
large. 
\citet{izotovetal06} attributed the large scatter to substantial  
flux errors of the weak [O\,{\sc ii}]$\lambda 7320+\lambda 7330$ emission 
lines. These investigations 
suggest that the O$^+$/H$^+$ abundances derived from 
the intensities of the auroral lines [OII]$\lambda$$\lambda$7320,7330, 
using Eq.(\ref{equation:o2i}), are somewhat uncertain.

\subsection{The (O/H)$_{\rm ff}$ abundances}

The spectra of high-metallicity 
H\,{\sc ii} regions generally exhibit strong [OII]$\lambda$3727 and  
[OIII]$\lambda$4959,5007 nebular lines, i.e. large R$_2$ and R$_3$ 
line intensity ratios. However,   
the [OIII]$\lambda$4363 auroral line is generally not detected, i.e. 
R cannot be measured directly. 
This prevents the use of the T$_{\rm e}$ method for abundance determination. 
However, \citet{ff,ohmax} have established 
a relation -- called the ff relation -- of the form R = f(R$_2$,R$_3$). 
Using this relation, the intensity 
of the auroral line can be estimated if measurements of the two nebular lines 
are available.

For the  H\,{\sc ii} regions in our sample, 
the R and R$_3$ ratios can be measured directly   
from the SDSS spectra.
But we have seen that 
for those galaxies with a redshift less than 0.024, the [OII]$\lambda$3727  
is out of the observed spectral range, and R$_2$ is not 
directly available.  
However, since there is relation of the type R = f(R$_2$,R$_3$),
there must also be a relation of the type R$_2$ = f(R,R$_3$), i.e. 
R$_2$ can be estimated if measurements of R and R$_3$  
are available.
Using the same 31 calibrating H\,{\sc ii} regions employed by \citet{ohmax} 
to establish their  R = f(R$_2$,R$_3$) relation,
we have derived the following relation
\begin{eqnarray}
\log R_{2,ff} & =  & 2.98 - 0.44 \log R_3 - 1.01 \, (\log R_3)^2  \nonumber  \\
               & +  & 1.37 \log R + 0.14 \, (\log R)^2 .
\label{equation:R2qo3}
\end{eqnarray}
Thus, we will use Eq.(\ref{equation:R2qo3}) to estimate R$_2$ and then  
determine the oxygen abundance 
with the help of Eqs. \ref{equation:o3} and (\ref{equation:o2}). 
The oxygen abundance derived in this way will be referred to 
as (O/H)$_{\rm ff}$.

\section{OXYGEN ABUNDANCES IN THE SPIRAL GALAXY M101: A COMPARISON   
OF O ABUNDANCES DERIVED BY DIFFERENT METHODS}

We focus here our attention on the  
Sc spiral galaxy M101 = NGC 5457. This galaxy  
has long served as the prototype system for studying radial oxygen 
abundance gradients in spiral disks. 
Many spectroscopic studies of H\,{\sc ii} 
regions in the disk of M101 have been carried out 
\citep{smith75,shieldssearle78,rayoetal82,mrs85,torrespeimbertetal89,
garnettkennicutt94,kinkelrosa94,kennicuttgarnett96,vanzeeetal98,garnettetal99,
luridianaetal02,kennicuttetal03,bresolin07}.
As a result, there are in the literature some 40 measurements of H\,{\sc ii} 
regions in the disk of M101 with a detected [OIII]$\lambda$4363 line, 
and the radial oxygen abundance gradient in the spiral galaxy is  
well established \citep{kennicuttetal03,m101,pvc04}.
This abundance of data allows us to test the reliability of the 
oxygen abundances derived from the SDSS spectra, using different methods.

\subsection{Comparison between the 1- and 2-dimensional 
t$_2$--t$_3$ relations}

The electron temperature $t_2$ is 
usually determined from a relation between $t_2$ to 
$t_3$, derived by fitting H\,{\sc ii} region photoionization models. 
Several versions of this $t_2$ -- $t_3$ relation have been proposed  
\citep{campbelletal86,garnett92,pageletal92,izotovetal97,deharvengetal00,oeyshields00}.
In particular, a one-dimensional model-independent $t_2$ -- $t_3$ relation
has been proposed by  \citet{tt1}. 
\begin{equation} 
t_2 =  0.72 \, t_3 + 0.26 
\label{equation:tt1}
\end{equation} 
Recently,  \citet{tt} has found evidence suggesting that,   
instead of a one-to-one correspondence between t$_2$ 
and t$_3$, the t$_2$ -- t$_3$ relation is also dependent on a second 
parameter, the excitation parameter P (defined in Section 1). 
\citet{tt} has derived a two-dimensional parametric relation of the form 
\begin{equation}
\frac{1}{t_2}  =    0.41 \, \times \frac{1}{t_3} 
- 0.34 \, \times P  +  0.81     .
\label{equation:ttp}
\end{equation}

How do the above 1- and 2-dimensional t$_2$ -- t$_3$ relations 
stand up to the available data for M101?
Using Eqs.(\ref{equation:o3}) and (\ref{equation:o2}) together 
with the parametric 
t$_2$ -- t$_3$ relation (Eq. (\ref{equation:ttp})), 
we have computed standard oxygen abundances based on published 
observations of H\,{\sc ii} regions in the disk of M101
\citep{smith75,shieldssearle78,rayoetal82,mrs85,torrespeimbertetal89,
garnettkennicutt94,kinkelrosa94,kennicuttgarnett96,vanzeeetal98,garnettetal99,
luridianaetal02,kennicuttetal03,bresolin07}. 
The derived (O/H)$_{\rm st}$ abundances are shown by open circles in panel 1a 
of Fig.\ref{figure:ngc5457}, as a function of galactocentric radius normalized 
to the disk isophotal radius R$_{25}$. The galactocentric distances of the 
H\,{\sc ii} regions have been taken from \citet{kennicuttgarnett96}. 
The solid line is the linear best fit to those data (42 points)
\begin{equation}
12+\log(O/H)  = 8.67 \, (\pm 0.03) \; - 0.71 \, (\pm 0.05) \, R_G/R_{25} .
\label{equation:ztenew}
\end{equation} 
The dotted lines show shifts of $\pm$ 0.1 dex from the best fit. 
The mean deviation of log(O/H)$_{\rm st}$ from the best fit is 0.073 dex.

We have then computed the standard oxygen abundances for the same sample
of H\,{\sc ii} region spectra, but using the 1-dimensional t$_2$ -- t$_3$ 
relation (Eq. (\ref{equation:tt1})). The resulting (O/H)$_{\rm st}$ abundances 
are shown by open circles as a function of galactocentric distance 
in panel 1b of Fig.\ref{figure:ngc5457}. 
Again, the solid line is the linear best fit to those data (42 points)
\begin{equation}
12+\log(O/H)  = 8.76 \, (\pm 0.04) \; - 0.85 \, (\pm 0.06) \, R_G/R_{25} 
\label{equation:zteold}
\end{equation} 
and the dotted lines are shifts of $\pm$ 0.1 dex from the best fit.  
The mean value of the deviations of log(O/H)$_{\rm st}$ from the best 
fit is 0.089 dex.
Comparison between Eqs. (\ref{equation:ztenew}) and (\ref{equation:zteold}) 
shows that the radial distributions of 
the standard oxygen abundances derived with the 1- and 2-dimensional
t$_2$ -- t$_3$ relations are rather similar. However, the scatter of the 
points in panel 1a obtained with the parametric relation is slightly lower 
than that in the diagram obtained with the one-dimensional relation. The 
agreement between the two diagrams is not surprising since the one-dimensional 
relation is just a kind of ``average'' of the parametric relation \citep{tt}. 
Therefore, while the differences between the oxygen abundances derived with 
those relations can be appreciable for individual H\,{\sc ii} regions depending 
on their excitation level, the mean difference for a sample of H\,{\sc ii} 
regions with different excitation parameters, varying in a random manner,  
should be near zero. We note that the one-dimensional relation derived by 
\citet{tt1} is close to the one proposed by \citet{campbelletal86} which has 
found wide acceptance and use. Consequently, the radial distributions of the 
standard oxygen abundances derived here with both the one-dimensional and 
parametric t$_2$ -- t$_3$ relations are similar to the one found  
for the disk of M101 in previous studies: 
\begin{equation}
12+\log(O/H)  = 8.76 \, (\pm 0.06) \; - 0.90 \, (\pm 0.08) \, R_G/R_{25} 
\label{equation:kbg}
\end{equation} 
\citep{kennicuttetal03} (Eq. (5)) and
\begin{equation}
12+\log(O/H)  = 8.79 \, (\pm 0.03) \; - 0.88 \, (\pm 0.08) \, R_G/R_{25} 
\label{equation:pvc}
\end{equation} 
\citep{pvc04} (Eq. (28)).

\subsection{Comparison of O abundances derived by 
various methods}

The [OIII]$\lambda$4363 auroral line is seen in 20 SDSS spectra of 
H\,{\sc ii} regions in the disk of M101 (Fig.\ref{figure:l4363}).
These spectra are listed in Table \ref{table:nearby}. 
The [OIII]$\lambda$7320+$\lambda$7330 nebular lines are measurable in 19 
of these spectra, allowing determination of (O/H)$_{7325}$.
The only exception is spectrum \# 1324-53088-236 (Fig.\ref{figure:l7325}).
There must be some overlap between the SDSS targets and those of previous 
spectroscopic studies of H\,{\sc ii} regions in the disk of M101. 
However, we have not attempted cross-identifications because of 
the poor accuracy of  
 the H\,{\sc ii} region positions used in previous studies. These positions 
generally come from the catalog of \citet{hodge90} and 
have uncertainties of 0.2-1 s of time in right ascension and 3-10 arcsec 
in declination. This means that two or more Hodge et al's
objects can be within an error box of 1s-10 arcsec
 centered on the SDSS position,
which may result in misidentifications.
  
\begin{figure}
\begin{center}
\resizebox{0.75\hsize}{!}{\includegraphics[angle=0]{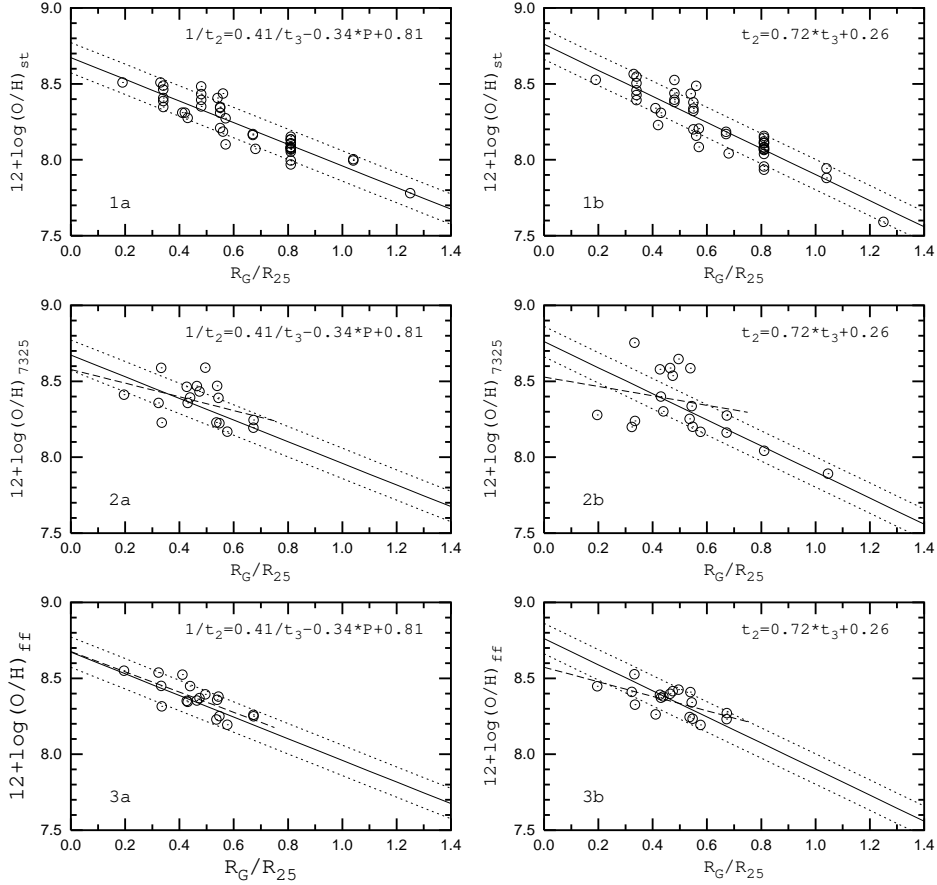}}
\caption{
{\footnotesize 
Radial distribution of the oxygen abundance in the spiral galaxy M101.
The abundances (plotted as open circles) in the left panels have been 
calculated 
using the two-dimensional parametric t$_2$ -- t$_3$ relation 
(Eq. (\ref{equation:ttp})), while those in the right panels have been 
calculated using the one-dimensional 
t$_2$ -- t$_3$ relation (Eq. (\ref{equation:tt1})).
In all panels, abundances are plotted as a function of 
galactocentric distance, normalized to the galactic disk isophotal radius
R$_{25}$. 
{\it Panels 1a and 1b.} 
Standard oxygen abundances (O/H)$_{st}$ are plotted.
The solid line is the linear least-squares best fit to those abundances.
 The dotted lines represent shifts of $\pm$ 0.1 dex from the best fit. 
{\it Panel 2a and 2b.} 
(O/H)$_{7325}$ abundances are plotted.  
The dashed line is the linear least-squares best fit to those abundances.
The solid and dotted lines in panel 2a 
are the same as in panel 1a, and those in panel 2b are the 
same as in panel 1b.
{\it Panel 3a and 3b.} 
(O/H)$_{ff}$ abundances are plotted. 
The dashed line is the linear least-squares best fit to those abundances.
The solid and dotted lines in panel 3a 
are the same as in panel 1a, and those in panel 3b are the 
same as in panel 1b.
}
}
\label{figure:ngc5457}
\end{center}
\end{figure}

\begin{figure}
\resizebox{0.80\hsize}{!}{\includegraphics[angle=0]{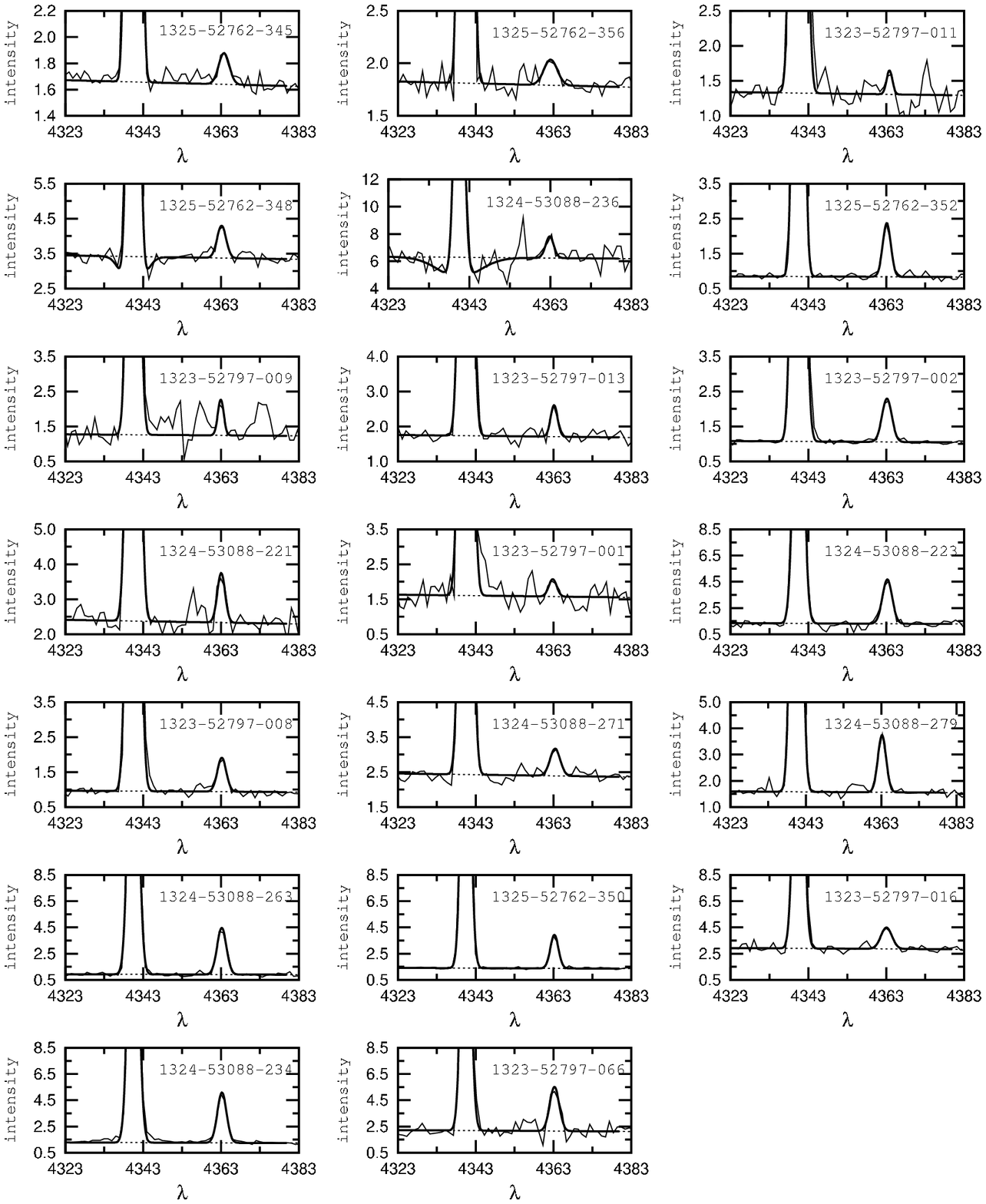}}
\caption{ 
Fits to the H${\gamma}$ and [OIII]$\lambda$4363 lines in spectra of  
H\,{\sc ii} regions in the spiral galaxy M101. 
The thin solid line is the observed spectrum, the thick solid line the fit
to it, and the dashed line the fit to the continuum.
}
\label{figure:l4363}
\end{figure}

\begin{figure}
\resizebox{0.80\hsize}{!}{\includegraphics[angle=0]{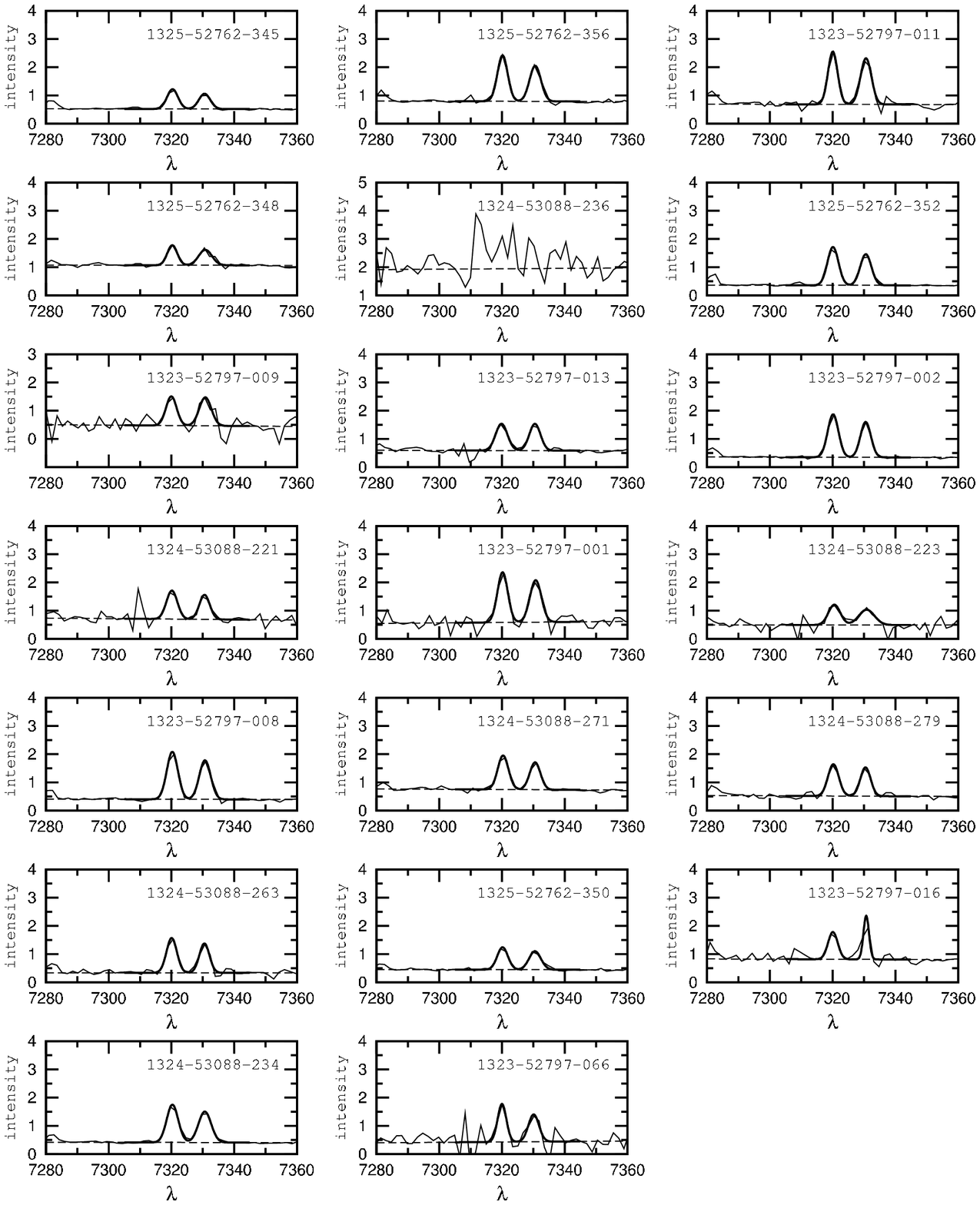}}
\caption{ 
Fits to the [OII]$\lambda$7320 and [OII]$\lambda$7330 lines in spectra of 
H\,{\sc ii} regions in the spiral galaxy M101. 
The thin solid line is the observed spectrum, the thick solid line the fit 
to it, and the dashed line 
the fit to the continuum.
}
\label{figure:l7325}
\end{figure}

The radial distribution of (O/H)$_{7325}$ abundances derived with the 
parametric t$_2$ -- t$_3$ relation is 
shown by open circles in panel 2a of Fig.\ref{figure:ngc5457}. 
The solid and dotted lines in panel 2a are the same as in panel 1a. 
The mean deviation of log(O/H)$_{7325}$ from the radial gradient traced by the 
standard abundances (O/H)$_{\rm st}$ (Eq. (\ref{equation:ztenew}))
is 0.124 dex. 
The dashed line is the linear best fit to the (O/H)$_{7325}$ data 
(17 data points)
\begin{equation}
12+\log(O/H)  = 8.58 \, (\pm 0.12) \; - 0.45 \, (\pm 0.25) \, R_G/R_{25} .
\label{equation:zrinew}
\end{equation} 
The galactocentric distances of the the H\,{\sc ii} regions 
have been computed from the coordinates associated with each spectrum 
and given in the SDSS database. The coordinates of the center of M101, 
its inclination and its position  
angle are from the Third Reference Catalog of Bright Galaxies 
\citep{rc3}. The temperature t$_2$ can be derived from the parametric 
relation only if R$_2$, and hence the excitation parameter P,  can 
be determined. 
Since R$_2$ cannot be estimated from Eq. (\ref{equation:R2qo3}) 
for low-metallicity H\,{\sc ii} regions, then t$_2$ cannot be
derived through the parametric t$_2$ -- t$_3$ relation. Hence, 
in panel 2a of Fig.\ref{figure:ngc5457}, 
(O/H)$_{7325}$ abundances have been derived only for high-metallicity 
H\,{\sc ii} regions.

The radial distribution of (O/H)$_{7325}$ abundances derived with the 
one-dimensional t$_2$ -- t$_3$ relation is 
shown by open circles in panel 2b of Fig.\ref{figure:ngc5457}. 
The solid and dotted lines in panel 2b are the same as in panel 1b.
The mean deviation of log(O/H)$_{7325}$ from the radial gradient traced by the 
standard abundances (O/H)$_{\rm st}$ (Eq. (\ref{equation:zteold})) 
is 0.195 dex. The dashed line is the linear best fit to the (O/H)$_{7325}$
 data (17 data points; the two low-metallicity H\,{\sc ii} regions at 
the largest galactocentric distances were excluded)
\begin{equation}
12+\log(O/H)  = 8.53 \, (\pm 0.18) \; - 0.31 \, (\pm 0.38) \, R_G/R_{25} .
\label{equation:zriold}
\end{equation} 

The radial distribution of (O/H)$_{\rm ff}$ abundances derived with the 
parametric t$_2$ -- t$_3$ relation is shown 
by open circles in panel 3a of Fig.\ref{figure:ngc5457}. 
The solid and dotted lines in panel 3a are the same as in panel 1a. 
The mean deviation of log(O/H)$_{\rm ff}$ from the radial 
gradient traced by the 
standard abundances (O/H)$_{\rm st}$ (Eq. \ref{equation:ztenew})) 
is 0.071 dex. 
The dashed line is the linear best fit to (O/H)$_{\rm ff}$ data (18 data points)
\begin{equation}
12+\log(O/H)  = 8.68 \, (\pm 0.07) \; - 0.67 \, (\pm 0.14) \, R_G/R_{25} .
\label{equation:zffnew}
\end{equation} 
The (O/H)$_{\rm ff}$ abundances have been derived from the 18 SDSS spectra 
with a detected [OIII]$\lambda$4363 auroral line. Two spectra, 
1324-53088-234 and 1323-52797-066, have been excluded because they pertain 
to H\,{\sc ii} regions at large galactocentric distances and hence  
with low-metallicity. Eq. (\ref{equation:R2qo3}) 
is not applicable in this regime, since it was derived 
only for H\,{\sc ii} regions with 12 +log (O/H) $\ga$ 8.2.

The radial distribution of (O/H)$_{\rm ff}$ abundances derived with the 
one-dimensional t$_2$ -- t$_3$ relation is shown 
by open circles in panel 3b of Fig.\ref{figure:ngc5457}. 
The solid and dotted lines in panel 3b are the same as in panel 1b. 
The mean deviation of log(O/H)$_{\rm ff}$ from the radial 
gradient traced by the 
standard abundances (O/H)$_{\rm st}$ (Eq. (\ref{equation:zteold})) 
is 0.082 dex. 
The dashed line is the linear best fit to (O/H)$_{\rm ff}$ data (18 data points)
\begin{equation}
12+\log(O/H)  = 8.57 \, (\pm 0.07) \; - 0.48 \, (\pm 0.14) \, R_G/R_{25} .
\label{equation:zffold}
\end{equation}

Comparison between panels 1a and 3a of Fig.\ref{figure:ngc5457} 
shows that the (O/H)$_{\rm ff}$ -- R$_{\rm G}$ and 
the (O/H)$_{\rm st}$ -- R$_{\rm G}$ diagrams derived with the parametric 
t$_2$ -- t$_3$ relation are in good agreement, in the sense that 
the mean deviation of the (O/H)$_{\rm ff}$ abundances (0.071 dex) 
from the radial trend traced by the (O/H)$_{\rm st}$ abundances is close to 
the mean deviation of the (O/H)$_{\rm st}$ abundances (0.073 dex). 
The global characteristics of the radial distributions (the central abundance 
and the slope) of the (O/H)$_{\rm ff}$ and (O/H)$_{\rm st}$ abundances 
are also close to each other (compare Eq. (\ref{equation:ztenew}) and 
Eq. (\ref{equation:zffnew})). 
It should be emphasized that the reliability of the determination 
of the global characteristics of the radial abundance distribution 
depends not only on the accuracy of the abundance determination in 
individual H\,{\sc ii} regions, but also on the range of
galactocentric distances over which these 
H\,{\sc ii} regions are distributed. 
Indeed, if we consider the (O/H)$_{\rm st}$ -R$_G$ diagram only for 
H\,{\sc ii} regions in the inner part of the disk, with galactocentric 
distances R$_G$/R$_{25}$ less than 0.6 (at this distance the value of 
12+log(O/H) decreases to $\approx$ 8.2, the lower limit for the 
applicability of our method for determining (O/H)$_{\rm ff}$ abundances) 
then the linear best fit to the resulting 23 data points is
\begin{equation}
12+\log(O/H)  = 8.63 \, (\pm 0.08) \; - 0.59 \, (\pm 0.17) \, R_G/R_{25} .
\label{equation:ztenr}
\end{equation} 
It is worth noting that the agreement between the central abundance and the 
slope derived from (O/H)$_{\rm st}$ abundances for H\,{\sc ii} regions in 
the inner part of the disk, Eq. (\ref{equation:ztenr}), 
and those quantities 
derived from the same (O/H)$_{\rm st}$ abundances, but for H\,{\sc ii} regions 
in the whole disk, Eq. (\ref{equation:ztenew}), 
is worse than the agreement between the central abundance 
and the slope derived from (O/H)$_{\rm ff}$ abundances for 
H\,{\sc ii} regions in the same inner part of the disk,  
Eq. (\ref{equation:zffnew}), and those quantities 
derived from (O/H)$_{\rm st}$ abundances for H\,{\sc ii} regions in 
the whole disk, Eq. (\ref{equation:ztenew}). 
 
Comparison between panels 1b and 3b of Fig.\ref{figure:ngc5457} 
shows that the (O/H)$_{\rm ff}$ -- R$_{\rm G}$ and 
the (O/H)$_{\rm st}$ -- R$_{\rm G}$ diagrams derived with the  
one-dimensional t$_2$ -- t$_3$ relation are in reasonable 
agreement, or at least they are not in conflict. 
The mean deviation of the (O/H)$_{\rm ff}$ abundances  
from the radial trend traced by the (O/H)$_{\rm st}$ abundances (0.082 dex) 
is close to 
the mean deviation of the (O/H)$_{\rm st}$ abundances (0.089 dex). 
However, there appears to be a discrepancy between  
the global characteristics of the radial distribution 
of the (O/H)$_{\rm ff}$ abundances and those of the 
(O/H)$_{\rm st}$ abundances: they do not 
agree within the formal uncertainties (compare 
Eq. (\ref{equation:zteold}) with Eq. (\ref{equation:zffold})). 
This suggests that there may exist a small systematic error in 
the (O/H)$_{\rm ff}$ abundances. 
However, if we again consider the (O/H)$_{\rm st}$ -R$_G$ diagram only for 
H\,{\sc ii} regions in the inner part of the disk, 
then the linear best fit to the resulting 23 data points is
\begin{equation}
12+\log(O/H)  = 8.71 \, (\pm 0.10) \; - 0.74 \, (\pm 0.22) \, R_G/R_{25} ,
\label{equation:zteor}
\end{equation} 
Now the global characteristics of the radial distributions of the 
(O/H)$_{\rm ff}$ and (O/H)$_{\rm st}$ abundances agree within the formal 
uncertainties (compare Eq. (\ref{equation:zteor}) with 
Eq. (\ref{equation:zffold})). 
Alternatively, if we add to our set of (O/H)$_{\rm ff}$ abundances 
the data point corresponding to 
the H\,{\sc ii} region with the largest galactocentric distance 
(R$_{\rm G}$/R$_{25}$ = 1.25) in the set of (O/H)$_{\rm st}$ abundances,
then the linear best fit to this "extended" data set (19 data points) is 
\begin{equation}
12+\log(O/H)  = 8.75 \, (\pm 0.05) \; - 0.82 \, (\pm 0.09) \, R_G/R_{25} .
\label{equation:zffext}
\end{equation} 
Again,  the global characteristics of the radial distributions 
of the (O/H)$_{\rm ff}$ and (O/H)$_{\rm st}$ abundances  
agree within the formal uncertainties (compare 
Eq. (\ref{equation:zffext}) with Eq. (\ref{equation:zffold})). 
The fact that the (O/H)$_{\rm ff}$ abundances, derived here with both 
the one-dimensional and parametric t$_2$ -- t$_3$ relations, give a 
general radial trend close  
to the one traced by the (O/H)$_{\rm st}$ abundances 
suggests that reasonably accurate (O/H)$_{\rm ff}$ oxygen abundances 
can be derived for 
H\,{\sc ii} regions with oxygen abundances in the range 
8.20 $\la$ 12 + log (O/H) $\la$ 8.55, when 
only [OIII]$\lambda$4959+$\lambda$5007 and [OIII]$\lambda$4363 
line intensity measurements are available.  
However, a small systematic error 
in the (O/H)$_{\rm ff}$ abundances derived with 
the one-dimensional t$_2$ -- t$_3$ relation cannot be ruled out. 

Comparison of panels 1a and 2a and of panels 1b and 2b 
in Fig.\ref{figure:ngc5457} shows that the radial distribution of the 
(O/H)$_{7325}$ abundances follows the same general trend traced by the 
(O/H)$_{\rm st}$ abundances. Examination of panels 1a, 2a, and 3a shows that 
the scatter in the (O/H)$_{7325}$ -- R$_{G}$ diagram is appreciably larger 
than the scatter in the (O/H)$_{\rm st}$ -- R$_{G}$ and
(O/H)$_{\rm ff}$ -R$_{G}$ diagrams. 
Comparison between panels 2a and 2b shows that 
the scatter in the (O/H)$_{7325}$ -- R$_{G}$ diagram is lower for 
the parametric t$_2$ -- t$_3$ relation than
the one-dimensional relation. 
This suggests that the parametric 
relation gives more accurate values of t$_2$ as compared 
to the one-dimensional relation. Fig.\ref{figure:ngc5457} also shows that 
using the parametric relation instead of 
the one-dimensional one results in a significant decrease 
of the scatter in the (O/H)$_{7325}$ -- R$_{G}$ diagram, but only
in a marginal one in the (O/H)$_{\rm st}$ -- R$_{G}$ and 
(O/H)$_{\rm ff}$ -- R$_{G}$ diagrams. This can be understood 
by the fact that the relation between the 
O$^+$/H$^+$ abundance and the [OII]$\lambda$7320+$\lambda$7330 
line intensity (Eq. (\ref{equation:o2})) 
has a stronger dependence on t$_2$ than the one  
between the O$^+$/H$^+$ abundance  and the [OII]$\lambda$3727 line intensity
(Eq. (\ref{equation:o2i})). 
Therefore, the same error in t$_2$ results in a larger error in 
the (O/H)$_{7325}$ abundances than in the (O/H)$_{\rm st}$ 
and (O/H)$_{\rm ff}$ abundances.

In summary, we have arrived at the following main conclusions: 
1) reasonably accurate (O/H)$_{\rm ff}$ abundances can be derived for 
H\,{\sc ii} regions with oxygen abundances in the range 
8.20 $\la$ 12 + log (O/H) $\la$ 8.55,  
when only [OIII]$\lambda$4959+$\lambda$5007 and [OIII]$\lambda$4363 line 
intensity measurements are available; 
2) there is a hint that the parametric t$_2$ -- t$_3$ relation provides 
more accurate values of t$_2$ than the one-dimensional relation.

\section{OXYGEN ABUNDANCES IN NEARBY GALAXIES}

\subsection{Abundances and abundance gradients}

We have extracted from the Data Release 5 of the 
SDSS spectra of individual H\,{\sc ii} regions in 
about a hundred nearby galaxies which have the particularity 
of possessing two or more 
spectra of the same or different H\,{\sc ii} regions with a 
detected [OIII]$\lambda$4363 line. 
We have determined the (O/H)$_{\rm ff}$  and (O/H)$_{7325}$ abundances 
for each H\,{\sc ii} regions. 
The final list, given in Table \ref{table:sample}, 
consists of 29 galaxies, with 84 individual oxygen 
abundance determinations.
The following criteria have been used to select the objects in the final list:
1) their H\,{\sc ii} regions have high metallicities, i.e.  
12 + log (O/H) $\ga$ 8.20, because  
Eq. (\ref{equation:R2qo3}) is only applicable in that metallicity range;  
2) (O/H)$_{\rm ff}$  abundances 
derived from two or more SDSS spectra of the same H\,{\sc ii} region,
when available, differ by less than 0.1 dex; 
and 3) the (O/H)$_{\rm ff}$ abundances of different H\,{\sc ii} regions 
within the same galaxy, but with similar galactocentric distances are 
consistent with each other, i.e. their abundances differ by less than 0.1 dex.
Criteria 2 and 3 are designed to select only 
galaxies with reliably determined oxygen abundances.
Those criteria were not applied to M101, in this case all the 
H\,{\sc ii} regions with measured [OIII]$\lambda$4363 line were included in 
the consideration. 

The measured line intensities and derived abundances for our 
sample of nearby galaxies are given in Table \ref{table:nearby}. 
For each galaxy, we give in column 2 
the SDSS spectrum number, composed of the 
plate number, the modified Julian date of observations and the  
number of the fiber on the plate.
The galactocentric distance normalized to the isophotal radius
of the galaxy is listed in column 3. 
The galactocentric distances were obtained
from the SDSS coordinates of the spectral observations. The coordinates of 
the center of each galaxy, its diameter, inclination 
and major axis position
angle were taken from the Lyon/Meudon Extragalactic Database (LEDA) and 
the NASA/IPAC Extragalactic Database (NED). 
The dereddened intensities and uncertainties of the 
[OIII]$\lambda$4363, [OIII]$\lambda$4959+$\lambda$5007 and  
[OII]$\lambda$7320+$\lambda$7330 lines are given respectively 
in columns 4, 5 and 6.
The line intensities have been scaled so that I$_{H\beta}$ = 1.
The electron temperature t$_3$ within the O$^{++}$ zone is shown  
in column 7. The (O/H)$_{7325}$ and (O/H)$_{\rm ff}$ abundances  
determined 
with the parametric relation (Eq. (\ref{equation:ttp})) are 
given respectively in columns 8 and 9, while  
the same abundances determined 
with the one-dimensional relation (Eq. (\ref{equation:tt1})) are 
given respectively in columns 10 and 11. 

Examination of Table \ref{table:nearby} shows that the 
vast majority of the H\,{\sc ii} regions 
in our sample have oxygen abundances in the 
very narrow interval 8.2 $\la$ 12 + log (O/H) $\la$ 8.4. 
This narrow metallicity range is due to two selections effects. 
As noted before, H\,{\sc ii} regions with 12 + log (O/H) $\la$ 8.2 
have been excluded because  
Eq. (\ref{equation:R2qo3}) is only 
applicable to high-metallicity H\,{\sc ii} regions. 
At the other metallicity end, 
the [OIII]$\lambda$4363 line is too weak to be detected in  
SDSS spectra of H\,{\sc ii} regions with 12+log(O/H) $\ga$ 8.4. 

\begin{figure}
\begin{center}
\resizebox{0.75\hsize}{!}{\includegraphics[angle=0]{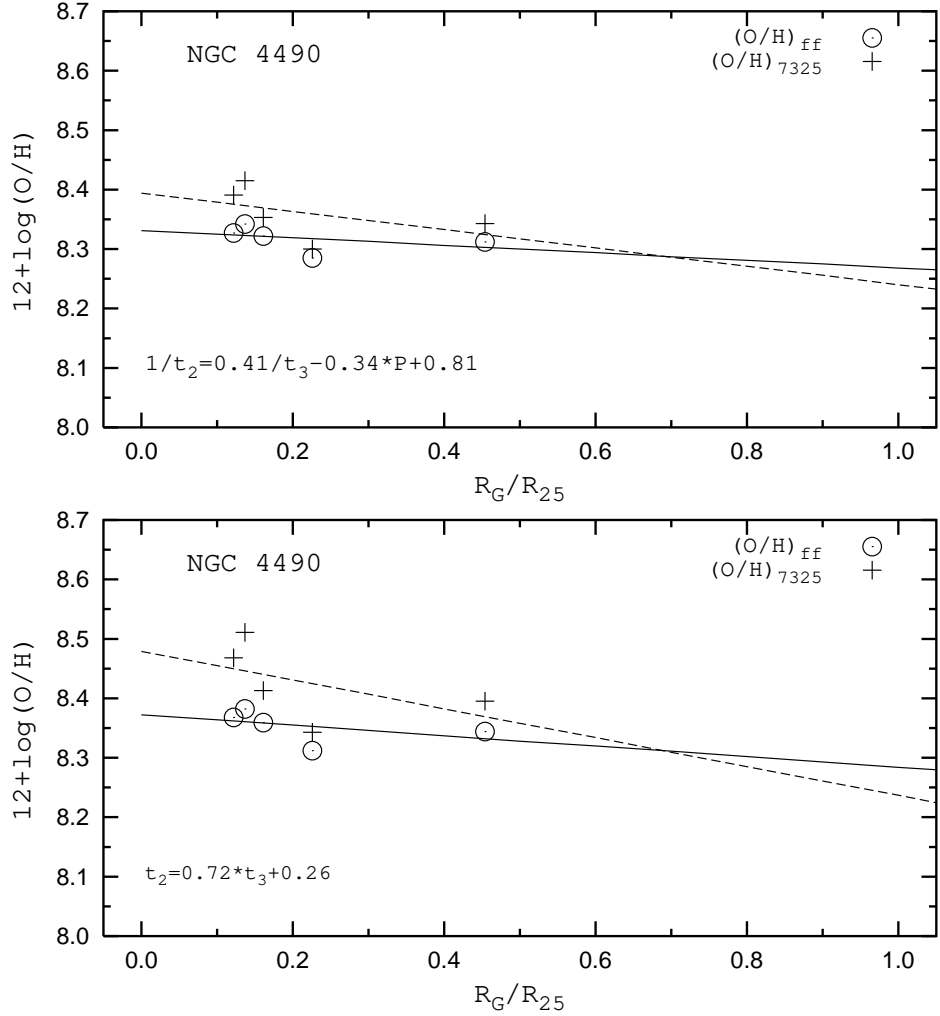}}
\caption{
Radial distribution of the oxygen abundance in the disk of the 
spiral galaxy NGC 4490. 
{\it Top panel.} 
(O/H)$_{\rm ff}$ (open circles) and (O/H)$_{7325}$ (plus signs) 
abundances as a function of galactocentric distance, normalized to the 
galactic disk isophotal radius R$_{25}$. The parametric 
t$_2$ -- t$_3$ relation (Eq. (\ref{equation:ttp})) has been used. 
The solid line is the linear least-squares fit to (O/H)$_{\rm ff}$ data, 
and the dashed line is the linear least-squares fit to (O/H)$_{7325}$ data. 
{\it Bottom panel.} 
The same as in the top panel, except that 
the one-dimensional t$_2$ -- t$_3$ relation 
(Eq. (\ref{equation:tt1})) has been used. 
}
\label{figure:ngc4490}
\end{center}
\end{figure}

In our sample, there are two galaxies (besides M101), which each have five 
distinct H\,{\sc ii} region measurements with a detected [OIII]$\lambda$4363 
line. The first galaxy is NGC 4490, and its measured H\,{\sc ii} regions
span a large enough range of galactocentric distances to allow  
a derivation of the metallicity radial distribution in its disk.  
Its (O/H)$_{\rm ff}$ and (O/H)$_{7325}$ abundances derived with the 
parametric relation are shown respectively by open circles and 
plus signs 
as a function of galactocentric distance in the top panel of 
Fig.\ref{figure:ngc4490}. 
 The solid line is the linear least-squares fit to the (O/H)$_{\rm ff}$  data: 
\begin{equation}
12+\log(O/H)  = 8.33 \, (\pm 0.02) \; - 0.063 \, (\pm 0.082) \, R_G/R_{25}
\label{equation:n4490ffnew}
\end{equation} 
The mean deviation is 0.017 dex.
 The dashed line is the linear least-squares fit to the (O/H)$_{7325}$  data: 
\begin{equation}
12+\log(O/H)  = 8.39 \, (\pm 0.04) \; - 0.154 \, (\pm 0.165) \, R_G/R_{25}
\label{equation:n4490rinew}
\end{equation} 
The mean deviation is 0.035 dex.

The bottom panel shows a similar diagram, 
but with the (O/H)$_{\rm ff}$ and (O/H)$_{7325}$ abundances derived with 
the one-dimensional relation.  
As in the top panel, the linear least-squares fit to the (O/H)$_{\rm ff}$ data 
is shown by the solid line:   
\begin{equation}
12+\log(O/H)  = 8.37 \, (\pm 0.03) \; - 0.088 \, (\pm 0.101) \, R_G/R_{25}
\label{equation:n4490ffold}
\end{equation}
The mean deviation is 0.021.
The linear least-squares fit to the (O/H)$_{7325}$ data 
is shown by the dashed line:   
\begin{equation}
12+\log(O/H)  = 8.48 \, (\pm 0.06) \; - 0.243 \, (\pm 0.237) \, R_G/R_{25}
\label{equation:n4490riold}
\end{equation}
The mean deviation is 0.050.

\begin{figure}
\begin{center}
\resizebox{0.75\hsize}{!}{\includegraphics[angle=0]{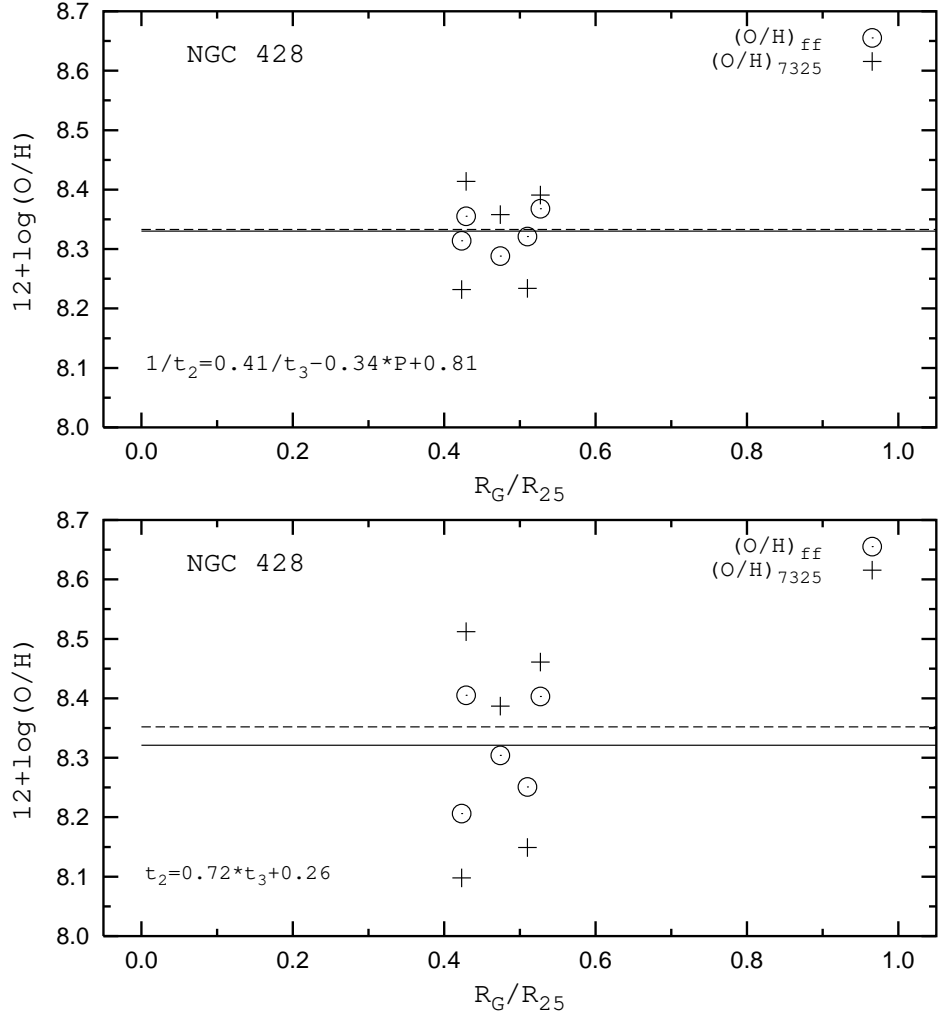}}
\caption{
Oxygen abundances in the disk of the magellanic irregular galaxy NGC 428.
{\it Top panel.} 
(O/H)$_{\rm ff}$ (open circles) and (O/H)$_{7325}$ (plus signs) 
abundances as a function of galactocentric distance, normalized 
to the galactic disk isophotal radius R$_{25}$. The parametric t$_2$ -- t$_3$ 
relation (Eq. (\ref{equation:ttp})) has been used. The solid line shows 
the mean (O/H)$_{\rm ff}$ abundance. 
The dashed line shows the mean (O/H)$_{7325}$ abundance. 
{\it Bottom panel.} 
The same as in the top panel, except that the one-dimensional t$_2$ -- t$_3$ 
relation (Eq. (\ref{equation:tt1})) has been used. 
}
\label{figure:ngc428}
\end{center}
\end{figure}

The second galaxy is NGC 428. As for NGC 4490, the top panel of 
Fig.\ref{figure:ngc428} shows the 
(O/H)$_{\rm ff}$ and (O/H)$_{7325}$ derived with the parametric 
relation as 
a function of galactocentric distance, while the 
bottom panel shows the same quantities derived with the one-dimensional 
relation. Unfortunately, the measured H\,{\sc ii} regions in 
NGC 428 cover a too small 
interval of galactocentric distances to allow a derivation of  
the radial distribution of oxygen abundances across the disk of the galaxy.
We can however estimate the mean value 
of the oxygen abundance within a small range of galactocentric distances.  
The mean values are: 12+log(O/H)$_{\rm ff}$ = 8.33 $\pm$ 0.03 and
12+log(O/H)$_{7325}$ = 8.33 $\pm$ 0.08, when determined with the parametric 
t$_2$ -- t$_3$ relation.
The mean values are: 12+log(O/H)$_{\rm ff}$ = 8.32 $\pm$ 0.08 and
12+log(O/H)$_{7325}$ = 8.35 $\pm$ 0.17,
 when determined with the one-dimensional 
t$_2$ -- t$_3$ relation.
As in our study of M101, we find that the scatter 
in (O/H)$_{\rm ff}$ abundances is smaller when determined with the parametric 
relation than when determined with the one-dimensional relation.   

The scatter also increases when going 
from (O/H)$_{\rm ff}$ to (O/H)$_{7325}$ abundances, even when both 
are determined with the same t$_2$ -- t$_3$ relation.
What is the origin of such an increase in scatter? 
From one point of view, we would expect the increase to be caused 
by measurement errors of the weak [OII]$\lambda$7320+$\lambda$7330 
emission lines. 
From another point of view, measurement errors 
of the [OIII]$\lambda$4363 emission line 
can also be responsible for the differences 
between (O/H)$_{\rm ff}$ and (O/H)$_{7325}$ abundances. 
Indeed, measurement errors of [OIII]$\lambda$4363 result in errors in the  
t$_3$ temperature which, in turn, induces errors in the t$_2$ temperature. 
As noted above,
 the relation between O$^+$/H$^+$ and [OII]$\lambda$7320+$\lambda$7330 
is more strongly dependent on t$_2$ than the 
relation between O$^+$/H$^+$ and [OII]$\lambda$3727
(compare  Eq. (\ref{equation:o2}) to Eq. (\ref{equation:o2i})). 
Therefore, the same error in t$_2$ results in a larger error in 
the (O/H)$_{7325}$ abundances than in the (O/H)$_{\rm st}$ 
and (O/H)$_{\rm ff}$ abundances which, in turn, causes differences 
between the (O/H)$_{\rm ff}$ and the (O/H)$_{7325}$ abundances. 

\begin{figure}
\begin{center}
\resizebox{0.50\hsize}{!}{\includegraphics[angle=0]{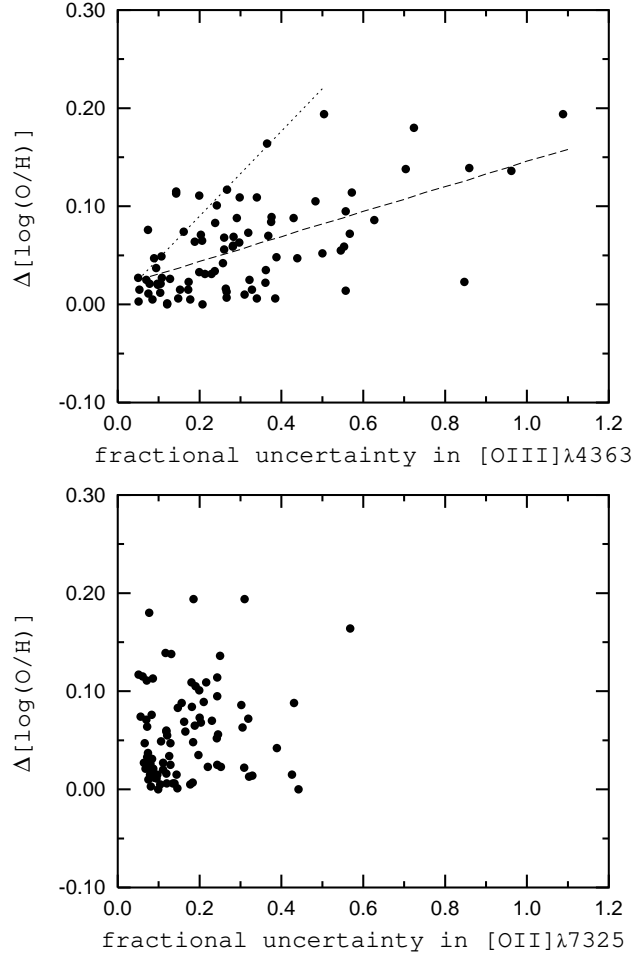}}
\caption{
{\footnotesize 
{\it Top panel.} 
Absolute value of the differences between 
log(O/H)$_{\rm ff}$ and log(O/H)$_{7325}$ 
abundances ($\Delta$log(O/H)=$\mid$log(O/H)$_{\rm ff}$-log(O/H)$_{7325}$$\mid$) 
versus the fractional uncertainty in the [OIII]$\lambda$4363 emission line.
Abundances have been 
determined with the parametric t$_2$ -- t$_3$ relation.
The dotted line shows an eye fit to the maximum values of the differences. The 
dashed line shows the linear least-squares fit to the data.
{\it Bottom panel.} 
Absolute value of the differences between 
log(O/H)$_{\rm ff}$ and log(O/H)$_{7325}$ abundances 
versus the fractional uncertainty 
in the [OII]$\lambda$7320+$\lambda$7330 emission lines. 
}
}
\label{figure:eedz}
\end{center}
\end{figure}

Thus measurement errors in [OIII]$\lambda$4363 and 
[OII]$\lambda$7320+$\lambda$7330 lines will both contribute  
to make the (O/H)$_{\rm ff}$ and 
the (O/H)$_{7325}$ abundances determined with the same t$_2$ -- t$_3$ 
relation differ from each other. 
Which contribution is greater? To answer that question, we have plotted
in  the top panel of Fig.\ref{figure:eedz} 
$\Delta$log(O/H) = $\mid$log(O/H)$_{\rm ff}$ -- log(O/H)$_{7325}$$\mid$,
where both abundances are determined with the parametric relation, against  
the fractional uncertainty in the 
measurement of the [OIII]$\lambda$4363 emission line 
(the uncertainty in the line measurement 
expressed in units of the line intensity).  
In the bottom panel of Fig.\ref{figure:eedz}, we have plotted  
the same quantity against the fractional uncertainty 
in the measurement of the [OII]$\lambda$7320+$\lambda$7330 emission lines. 
Inspection of the top panel of Fig.\ref{figure:eedz} shows that
the differences between  
(O/H)$_{\rm ff}$ and (O/H)$_{7325}$ are small  
in H\,{\sc ii} regions with an accurate measurement of the
  [OIII]$\lambda$4363 emission 
line, with an uncertainy of less than $\sim$ 10\%. 
The maximum difference  
between (O/H)$_{\rm ff}$ and (O/H)$_{7325}$ increases with increasing 
uncertainty in [OIII]$\lambda$4363. The dotted line 
shows an eye fit to the maximum differences. However, the correlation 
between $\Delta$(O/H) and the [OIII]$\lambda$4363 uncertainties is weak  
(dashed line). 
On the other hand, the H\,{\sc ii} regions that do have accurate   
[OII]$\lambda$7320+$\lambda$7330 measurements, 
with an uncertainy less than 10\%, still show large differences  
between (O/H)$_{\rm ff}$ and (O/H)$_{7325}$
(bottom panel of Fig.\ref{figure:eedz}). 
This suggests that the differences between (O/H)$_{\rm ff}$ and 
(O/H)$_{7325}$ abundances in our H\,{\sc ii} region sample  
are mainly due to measurement uncertainties 
in the [OIII]$\lambda$4363 line, and hence, to uncertainties
 in the electron temperature determination.

\subsection{Comparison of derived O abundances with previous work}

\subsubsection{The t$_2$ -- t$_3$ relation}

\begin{figure}
\begin{center}
\resizebox{0.50\hsize}{!}{\includegraphics[angle=0]{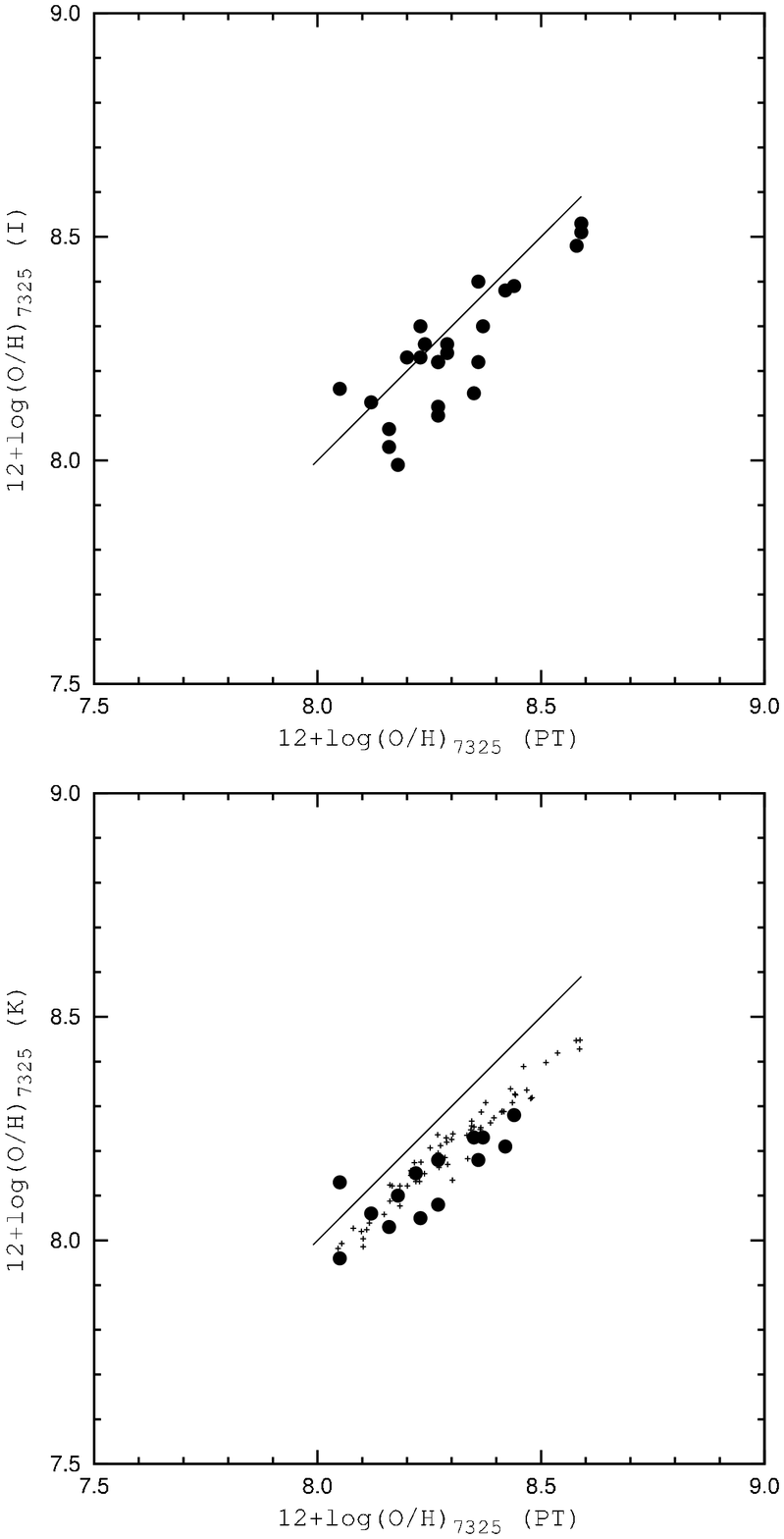}}
\caption{
{\footnotesize 
Comparison between our (O/H)$_{7325}$ abundances (labeled PT) derived 
with the one-dimensional t$_2$ -- t$_3$ relation,
and those determined by other authors, based on the same SDSS spectra.
{\it Top panel.} 
Abundances from \citet{izotovetal06} (labeled I) versus PT 
abundances (filled circles). 
The solid line corresponds to equality.
{\it Bottom panel.} 
Abundances from \citet{kniazevetal04} (labeled K) versus PT
abundances (filled circles). 
The solid line corresponds to equality. The plus
signs show (O/H)$_{7325}$ abundances derived with 
the t$_2$ -- t$_3$ relation used by \citet{kniazevetal04} as compared to  
abundances derived with our one-dimensional 
t$_2$ -- t$_3$ relation.
}
}
\label{figure:compar}
\end{center}
\end{figure}

Some H\,{\sc ii} regions in our sample have independent oxygen abundance 
determinations by other authors \citep{kniazevetal04,izotovetal06}, based on  
the same SDSS spectra. They 
 are listed in Table \ref{table:compar}.
In Fig.\ref{figure:compar}, we compare our (O/H)$_{7325}$ abundances 
(labeled PT on the x-axis), derived with the one-dimensional 
t$_2$ -- t$_3$ relation, with those obtained by \citet{izotovetal06}
(labeled I on the y-axis) in 
the top panel, and with those derived by \citet{kniazevetal04} 
(labeled K on the y-axis)  in the 
lower panel. The filled circles are individual H\,{\sc ii} regions from 
Table \ref{table:compar}, and the solid lines correspond to equal abundances.

\begin{figure}
\begin{center}
\resizebox{1.00\hsize}{!}{\includegraphics[angle=0]{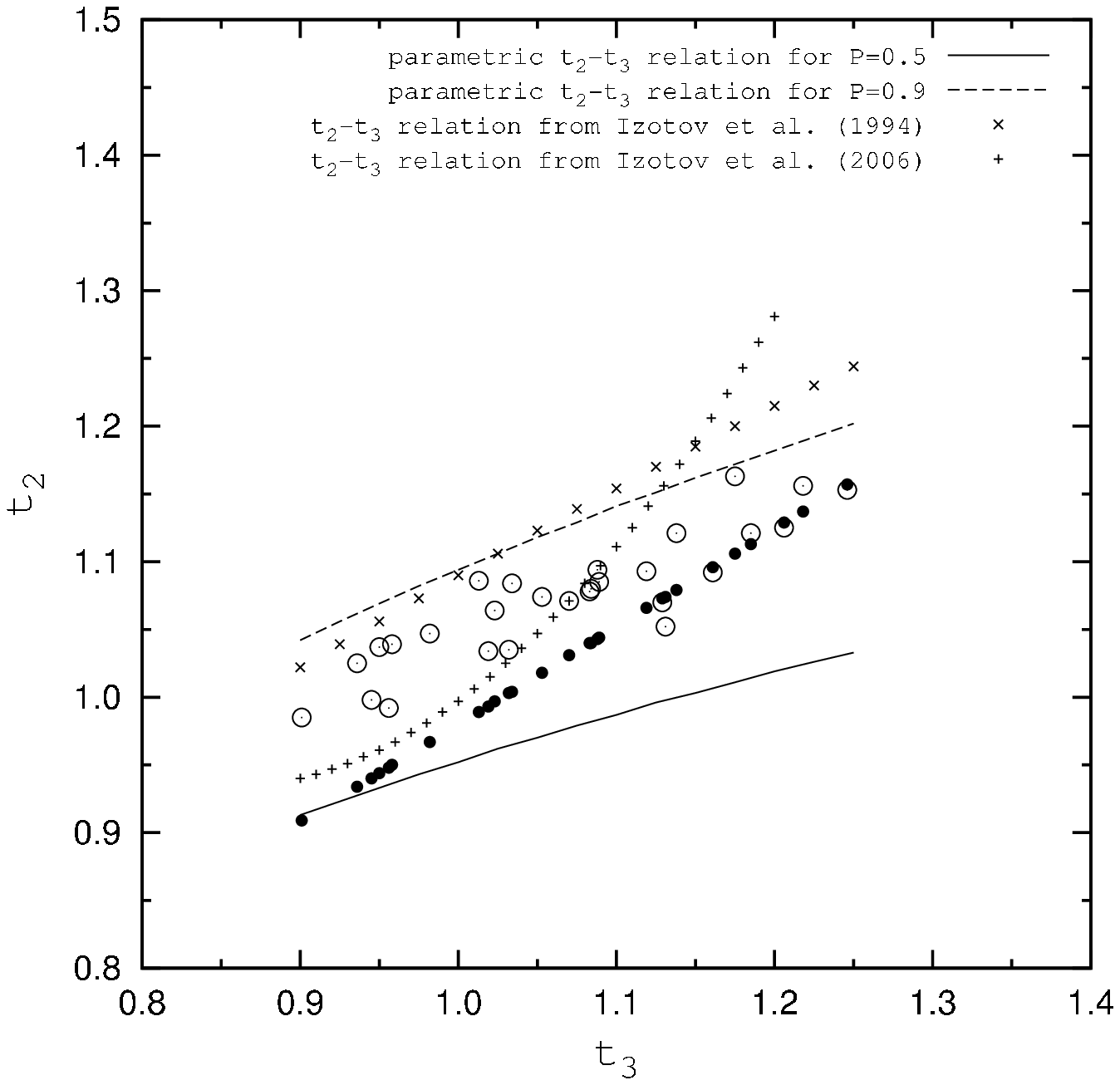}}
\caption{
The t$_2$ -- t$_3$ diagram. Open and filled 
circles denote H\,{\sc ii} regions with 
t$_2$ derived respectively with the two-dimensional parametric and 
one-dimensional t$_2$ -- t$_3$ relations. 
The curves corresponding to the parametric relation are shown 
for two different values of the excitation parameter, P = 0.5 (solid 
line) and P = 0.9 (dashed line). The t$_2$ -- t$_3$ relations 
from \citet{izotovetal94} and \citet{izotovetal06} 
are shown respectively by crosses and plus signs. 
}
\label{figure:tt}
\end{center}
\end{figure}

Inspection of Fig.\ref{figure:compar} 
and examination of Table \ref{table:compar} 
show that there are systematic differences between 
our (O/H)$_{7325}$ abundances and those 
obtained by the previously quoted authors, in 
the sense that our abundances are generally larger than theirs. 
The differences are especially large for the Kniazev et al.'s abundances
(bottom diagram). 
Since the same SDSS spectra are used in all cases, 
the differences must be due to differences in  
abundance determination methods, 
in particular in the t$_2$ -- t$_3$ relation used. 
Fig.\ref{figure:tt} shows the t$_2$ -- t$_3$ relations used by us and the 
above authors. Open and filled circles denote H\,{\sc ii} regions with 
[OIII]$\lambda$4363 and [OII]$\lambda$7320+$\lambda$7330 line measurements 
with uncertainties less than 20\%,
 with the t$_2$ temperature derived respectively 
from the parametric and one-dimensional t$_2$ -- t$_3$ relation.
The solid and dashed lines show 
the parametric relations respectively for two different 
values of the excitation parameter, P = 0.5 and 0.9.
 The t$_2$ -- t$_3$ relation 
of \citet{izotovetal94} used by \citet{kniazevetal04} is shown by crosses,
while the one used by \citet{izotovetal06} is shown by plus signs.
Fig.\ref{figure:tt} shows that there is a significant shift of the t$_2$ 
temperatures derived with our one-dimensional t$_2$ -- t$_3$ relation 
(filled circles) 
as compared to those derived by \citet{kniazevetal04} (crosses). 
It is this shift which is responsible for the differences 
between our and \citet{kniazevetal04} (O/H)$_{7325}$ abundances,
To illustrate this point directly, we have recomputed oxygen abundances 
for our H\,{\sc ii} region subsample 
with the t$_2$ -- t$_3$ relation used by \citet{kniazevetal04}.
These abundances are plotted against our 
abundances computed with the one-dimensional relation as 
small plus signs in the bottom panel of Fig.\ref{figure:compar}. 
The plus signs and the filled circles do indeed occupy the same general 
region. 

\subsubsection{The luminosity -- metallicity relation} 

It is interesting to 
check how our derived oxygen abundances fit in the luminosity -- central 
metallicity relation for galaxies, such as the one constructed by 
\citet{plato07}.  Since for the majority of the galaxies 
in our sample, the abundances were derived at only two galactocentric 
distances, the central metallicity can only be estimated for a handful  
of galaxies. 
We can estimate central metallicities for galaxies which satisfy  
one the following conditions: 1) the galaxy has  
at least one H\,{\sc ii} region at a galactocentric 
distance less than one tenth of the isophotal radius; or 2) 
it has a low-luminosity (M$_{\rm B}$ $\ga$ -18) and a morphological 
type Sm or later (i.e. the morphological type code given in Table 1 is 
higher than 8.0).  
It is known that the O/H distribution across the body of  
low-mass dwarf irregular galaxies does not show a gradient  
\citep{pageletal78,devostetal97}.
Then, the abundance in a H\,{\sc ii} region at any 
galactocentric distance is representative of
 the chemical composition of the dwarf galaxy's 
interstellar medium as a whole. 
 The galaxies so selected and 
for which a central metallicity can be 
estimated are listed in Table \ref{table:lummet}.  The central oxygen 
abundance derived in the cases of both the 
parametric and one-dimensional t$_2$ -- t$_3$ relations 
are shown for each galaxy respectively in columns 3 and 4. The 
central abundance, derived by extrapolating the linear fit in Fig. 5 
to R = 0, is given for the galaxy NGC 4490, while the average value of the  
oxygen abundance is given for the other galaxies.

\begin{figure}
\resizebox{0.75\hsize}{!}{\includegraphics[angle=0]{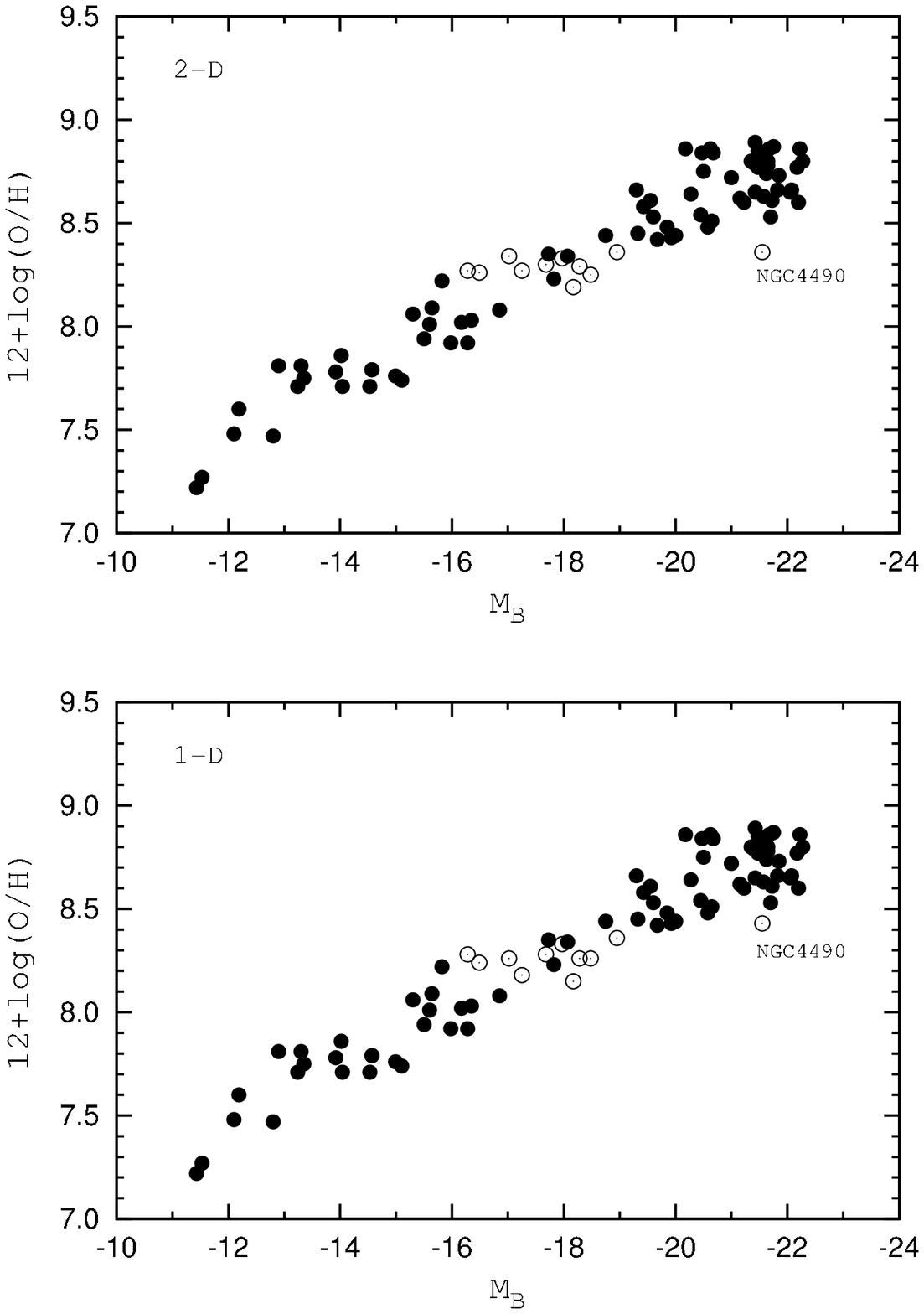}}
\caption{ 
The luminosity -- central metallicity diagram. 
{\it Top panel.} 
The open circles show the   
galaxies in Table \ref{table:lummet}, with abundances determined 
with the parametric t$_2$ -- t$_3$ relation. The filled circles show data 
from \citet{plato07}. 
{\it The bottom panel.} 
The same as in the top panel, except the abundances of the 
galaxies in Table \ref{table:lummet}  
have been determined 
with the one-dimensional t$_2$ -- t$_3$ relation. 
}
\label{figure:lz}
\end{figure}

The luminosity -- central metallicity diagram is shown in 
Fig.\ref{figure:lz}. 
The open circles denote the galaxies in our sample 
(Table \ref{table:lummet}) with abundances determined with the 
parametric t$_2$ -- t$_3$ relation in the upper panel, and 
with the one-dimensional relation in the bottom panel. 
The filled circles in both panels represent data from \citet{plato07}. 
Inspection of Fig.\ref{figure:lz} shows that the majority of the galaxies 
in our sample follow well the general trend of the luminosity -- central 
metallicity relation. This can be considered as an indirect evidence in 
favour of the reliability of our abundance determinations. 
An exception is the spiral galaxy NGC 4490 which has an oxygen 
abundance too low for its luminosity. This galaxy is 
the only interacting galaxy among all 
galaxies listed in Table \ref{table:lummet}. 
This suggests that metal-poor gas infall may be responsible for its 
relatively low metallicity.

\section{CONCLUSIONS}

We have determined oxygen abundances for a 
sample of nearby galaxies in the Data Release 5 
of the Sloan Digital Sky Survey, selected to 
possess two or more independent spectra of one or several H\,{\sc ii} 
regions with a detected [O\,{\sc iii}]~$\lambda 4363$ emission line. 
Since the nebular [OII]$\lambda$3727 line is out of the observed 
wavelength range in SDSS 
spectra of H\,{\sc ii} regions in nearby galaxies, we propose a method to 
determine (O/H)$_{\rm ff}$ oxygen abundances, 
using the classic T$_{\rm e}$ method coupled 
with the ff relation. 
Following \citet{kniazevetal04}, we have also determined (O/H)$_{7325}$
abundances, based on the 
intensities of the [OII]$\lambda$$\lambda$7320,7330 auroral lines,
and using  a small modification of the classic  
T$_{\rm e}$ method. 
Both (O/H)$_{\rm ff}$ and (O/H)$_{7325}$ abundances have been   
derived with one- and two-dimensional t$_2$ -- t$_3$ relations.

It was found that reliable (O/H)$_{\rm ff}$ abundances can be derived 
with measurements of only [OIII]$\lambda$4959+$\lambda$5007 and 
[OIII]$\lambda$4363 line intensities.  
The parametric two-dimensional t$_2$ -- t$_3$ relation appears to 
provide more accurate values of t$_2$ as compared to the  one-dimensional 
relation.
We show that the differences between (O/H)$_{\rm ff}$ and 
(O/H)$_{7325}$ abundances for our sample of H\,{\sc ii} regions 
are caused more by measurement uncertainties of the [OIII]$\lambda$4363 
line intensity  
than by those of the [OII]$\lambda$7320+$\lambda$7330 line intensity.   

Our final list contains 29 galaxies, with 84 separate determinations 
of H\,{\sc ii} region oxygen abundance.
Because of our selection methods, the metallicity of our galaxies lies in 
the narrow range 8.2 $\la$ 12 + log (O/H) $\la$ 8.4.  
The radial distribution of oxygen abundances in the disk of the spiral 
galaxy NGC 4490 is determined for the first time.
The objects in our sample generally follow the 
luminosity -- metallicity relation established for galaxies.
 An exception is NGC 4490  
which has an oxygen abundance lower than 
that of other galaxies of similar luminosities.  

\bigskip

    T.X.T. and L.S.P. acknowledge the support of National Science Foundation 
grant AST02-05785.  
    L.S.P. thanks the hospitality of the Astronomy Department of the 
University of Virginia where this investigation was carried out. 
    We thank the referee for the recommendations which improve the clarity 
of the presentation.
    We thank Yuri Izotov for useful discussions
    This research has made use of the NASA/IPAC Extragalactic Database 
(NED) which is operated by the Jet Propulsion Laboratory, California
Institute of Technology, under contract with the National Aeronautic and 
Space Administration.
    We acknowledge the work of the SDSS team. 
Funding for the SDSS has been provided by the
Alfred P. Sloan Foundation, the Participating Institutions, the National
Aeronautics and Space Administration, the National Science Foundation, the
U.S. Department of Energy, the Japanese Monbukagakusho, and the Max Planck
Society.

\clearpage


\begin{deluxetable}{lccccccl}
\rotate
\tabletypesize{\scriptsize}
\tablewidth{0pc}
\tablecaption{Nearby galaxy sample\label{table:sample}}
\tablehead{
galaxy                                                   & 
coordinates                                              & 
morphological                                            & 
morphological                                            & 
apparent B                                               & 
redshift                                                 & 
absolute B                                               & 
other names                                              \\
name\tablenotemark{a}                                    & 
                                                         & 
type                                                     & 
type code                                                & 
magnitude                                                & 
                                                         & 
magnitude                                                & 
                                                         }
\startdata 
M101                   &  14$^h$03$^m$12.44$^s$ + 54$^o$20$^{'}$53.0$^{''}$  &  SABc  &  5.9$\pm$0.3  &  8.39  &  0.000804  & -20.84  &  M102, NGC5457, UGC08981,\\
                       &                                                     &        &               &        &            &         &  ARP026, VV344, VV456,   \\  
                       &                                                     &        &               &        &            &         &  PGC050063               \\  
NGC428                 &  01$^h$12$^m$55.74$^s$ + 00$^o$58$^{'}$53.3$^{''}$  &  SABm  &  8.6$\pm$1.0  & 11.94  &  0.003843  & -19.48  &  UGC00763, UM309         \\
                       &                                                     &        &               &        &            &         &  PGC004367               \\  
NGC450                 &  01$^h$15$^m$30.55$^s$ - 00$^o$51$^{'}$39.2$^{''}$  &  SABc  &  5.9$\pm$0.5  & 12.72  &  0.005874  & -19.69  &  UGC00806, PGC004540     \\
NGC1110                &  02$^h$49$^m$09.37$^s$ - 07$^o$50$^{'}$19.5$^{''}$  &  SBm   &  8.8$\pm$1.0  & 14.58  &  0.004446  & -18.17  &  UGCA043, PGC010673      \\
NGC2541                &  08$^h$14$^m$40.06$^s$ + 49$^o$03$^{'}$41.7$^{''}$  &  SABc  &  6.0$\pm$0.4  & 12.18  &  0.001828  & -18.67  &  UGC04284, PGC023110     \\
NGC2552                &  08$^h$19$^m$19.54$^s$ + 50$^o$00$^{'}$28.5$^{''}$  &  SABm  &  9.0$\pm$0.5  & 12.68  &  0.001748  & -17.97  &  UGC04325, PGC023340     \\
NGC3023                &  09$^h$49$^m$52.70$^s$ + 00$^o$37$^{'}$04.1$^{''}$  &  SABc  &  5.4$\pm$0.8  & 13.39  &  0.006268  & -19.47  &  UGC05269, VV620,        \\
                       &                                                     &        &               &        &            &         &  PGC028272               \\  
NGC3319                &  10$^h$39$^m$09.51$^s$ + 41$^o$41$^{'}$13.0$^{''}$  &  SBc   &  5.9$\pm$0.4  & 11.72  &  0.002465  & -19.43  &  UGC05789, PGC031671     \\
NGC3432                &  10$^h$52$^m$30.93$^s$ + 36$^o$37$^{'}$09.3$^{''}$  &  SBm   &  8.9$\pm$0.5  & 11.64  &  0.002055  & -19.81  &  UGC05986, ARP206,       \\
                       &                                                     &        &               &        &            &         &  PGC032643, VV011        \\  
NGC3991                &  11$^h$57$^m$30.48$^s$ + 32$^o$20$^{'}$00.3$^{''}$  &  IB    &  9.7$\pm$1.3  & 13.52  &  0.010647  & -20.99  &  UGC06933, PGC037613,    \\
                       &                                                     &        &               &        &            &         &  VV523, CG0137           \\  
NGC3995                &  11$^h$57$^m$44.15$^s$ + 32$^o$17$^{'}$39.7$^{''}$  &  SABm  &  8.8$\pm$0.8  & 12.73  &  0.010854  & -21.55  &  UGC06944, PGC037624,    \\
                       &                                                     &        &               &        &            &         &  VV249, ARP313, CG0139   \\  
NGC4490                &  12$^h$30$^m$36.37$^s$ + 41$^o$38$^{'}$37.2$^{''}$  &  SBcd  &  7.0$\pm$0.2  &  9.77  &  0.001885  & -21.52  &  UGC07651, ARP269,       \\
                       &                                                     &        &               &        &            &         &  PGC041333, VV030        \\
NGC5669                &  14$^h$32$^m$43.92$^s$ + 09$^o$53$^{'}$28.2$^{''}$  &  SABc  &  6.0$\pm$0.4  & 12.66  &  0.004547  & -19.41  &  UGC09353, PGC051973     \\
IC2828                 &  11$^h$27$^m$11.03$^s$ + 08$^o$43$^{'}$53.1$^{''}$  &        &               & 14.70  &  0.003466  & -16.49  &  PGC035225, Tol1124+090  \\
UGC05189               &  09$^h$42$^m$56.75$^s$ + 09$^o$28$^{'}$18.3$^{''}$  &  I     &  9.9$\pm$0.5  & 14.62  &  0.010720  & -19.34  &  PGC027784, VV547        \\
UGC05249               &  09$^h$47$^m$45.37$^s$ + 02$^o$37$^{'}$38.6$^{''}$  &  Scd   &  6.8$\pm$0.8  & 14.50  &  0.006252  & -18.95  &  PGC028148               \\
UGC06596               &  11$^h$37$^m$51.06$^s$ + 56$^o$08$^{'}$32.7$^{''}$  &  IB    &  9.9$\pm$0.5  & 17.12  &  0.007609  & -16.28  &  PGC036000, VV148        \\
UGC09979               &  15$^h$42$^m$19.34$^s$ + 00$^o$28$^{'}$31.0$^{''}$  &  IB    &  9.8$\pm$1.0  & 14.93  &  0.006541  & -18.56  &  PGC055833, DDO201       \\
UGCA154                &  09$^h$16$^m$45.52$^s$ + 53$^o$26$^{'}$34.0$^{''}$  &  S?    &  3.2$\pm$5.0  & 14.71  &  0.007456  & -18.48  &  PGC026188, MRK0104,     \\
                       &                                                     &        &               &        &            &         &  SBS0913+536             \\
UGCA322                &  13$^h$04$^m$31.19$^s$ - 03$^o$34$^{'}$20.6$^{''}$  &  Sd    &  7.6$\pm$1.0  & 14.51  &  0.004536  & -17.31  &  PGC045195               \\
PGC001586              &  00$^h$25$^m$19.93$^s$ + 00$^o$31$^{'}$31.1$^{''}$  &  S?    &  4.6$\pm$5.0  & 16.61  &  0.014045  & -17.68  &  HS0022+0014, UM241      \\
PGC023706              &  08$^h$27$^m$18.05$^s$ + 46$^o$02$^{'}$02.8$^{''}$  &  S?    &  3.9$\pm$5.0  & 15.80  &  0.007344  & -17.25  &                          \\
PGC051971              &  14$^h$32$^m$45.11$^s$ + 02$^o$54$^{'}$54.2$^{''}$  &  Sbc   &  4.2$\pm$0.8  & 15.12  &  0.005070  & -17.02  &  Tol1430+031             \\
PGC056006              &  15$^h$46$^m$30.73$^s$ + 45$^o$59$^{'}$53.9$^{''}$  &        &               & 14.88  &  0.008878  & -18.28  &  MRK0490                 \\
HS1103+4346            &  11$^h$06$^m$29.03$^s$ + 43$^o$30$^{'}$41.3$^{''}$  &  BCG   &               &        &  0.021401  &         &  CG1377                  \\
HS1132+4416            &  11$^h$34$^m$53.68$^s$ + 44$^o$00$^{'}$16.4$^{''}$  &  BCG   &               &        &  0.018481  &         &                          \\
UM330                  &  01$^h$30$^m$03.62$^s$ + 00$^o$44$^{'}$34.9$^{''}$  &        &               &        &  0.017000  &         &                          \\
CG1419                 &  11$^h$23$^m$17.23$^s$ + 41$^o$03$^{'}$37.2$^{''}$  &        &               &        &  0.010529  &         &                          \\
SDSS125446.33+153529.8 &  12$^h$54$^m$46.32$^s$ + 15$^o$35$^{'}$29.9$^{''}$  &        &               &        &  0.008792  &         &                          \\
\enddata
\tablenotetext{a}{
The galaxies are listed in order 
of name category, with the following categories in descending order:  \\
M -- Messier catalogue of nebulae,  \\
NGC -- New General Catalogue,       \\
IC -- Index Catalogue,     \\
UGC -- Uppsala General Catalog of Galaxies,  \\
UGCA -- Uppsala General Catalog Appendix,  \\
PGC -- Principal Galaxy Catalog,  \\
HS -- Hamburg/ESO QSO Survey,  \\
UM -- University of Michigan Emission Line Objects,  \\
CG - Compact Group, \\
SDSS - Sloan Digital Sky Survey. 
}
\end{deluxetable}


\clearpage


\begin{deluxetable}{lrccccccccc}
\rotate
\tabletypesize{\scriptsize}
\tablewidth{0pc}
\tablecaption{Line intensities and derived oxygen abundances of H\,{\sc ii} 
regions in galaxies of nearby sample\label{table:nearby}}
\tablehead{
galaxy                                                   & 
spectrum                                                 & 
R$_{G}$\tablenotemark{a}                                 & 
[OIII]\tablenotemark{b}                                  & 
[OIII]\tablenotemark{b}                                  & 
[OII]\tablenotemark{b}                                   & 
t$_3$\tablenotemark{c}                                   & 
(O/H)$_{7325}$\tablenotemark{d}                          & 
(O/H)$_{ff}$\tablenotemark{d}                            &     
(O/H)$_{7325}$\tablenotemark{e}                          & 
(O/H)$_{ff}$\tablenotemark{e}                            \\     
name                                                     & 
number                                                   & 
                                                         & 
$\lambda$4363                                            & 
$\lambda$,$\lambda$4959,5007                             & 
$\lambda$7325                                            & 
                                                         & 
                                                         & 
                                                         & 
                                                         & 
}
\startdata 
                        &                &       &                     &                   &                   &       &       &       &       &        \\  
M101                   & 1325 52762 345 &  0.20 &  0.0045$\pm$ 0.0032 &   1.40$\pm$  0.01 &  0.031$\pm$ 0.004 &  0.90 &  8.41 &  8.55 &  8.28 &  8.45  \\
M101                   & 1325 52762 356 &  0.32 &  0.0072$\pm$ 0.0052 &   1.74$\pm$  0.01 &  0.033$\pm$ 0.003 &  0.96 &  8.36 &  8.54 &  8.20 &  8.41  \\
M101                   & 1323 52797 011 &  0.33 &  0.0038$\pm$ 0.0033 &   2.30$\pm$  0.02 &  0.039$\pm$ 0.005 &  0.76 &  8.59 &  8.45 &  8.75 &  8.53  \\
M101                   & 1325 52762 348 &  0.34 &  0.0178$\pm$ 0.0052 &   3.79$\pm$  0.02 &  0.026$\pm$ 0.004 &  1.00 &  8.23 &  8.32 &  8.24 &  8.33  \\
M101                   & 1324 53088 236 &  0.41 &  0.0272$\pm$ 0.0188 &   2.96$\pm$  0.04 &  0.000$\pm$ 0.000 &  1.24 &   ... &  8.52 &   ... &  8.26  \\
M101                   & 1325 52762 352 &  0.43 &  0.0220$\pm$ 0.0031 &   5.52$\pm$  0.07 &  0.060$\pm$ 0.005 &  0.95 &  8.46 &  8.35 &  8.58 &  8.39  \\
M101                   & 1323 52797 009 &  0.43 &  0.0126$\pm$ 0.0070 &   3.34$\pm$  0.04 &  0.038$\pm$ 0.012 &  0.94 &  8.36 &  8.34 &  8.40 &  8.37  \\
M101                   & 1323 52797 013 &  0.44 &  0.0097$\pm$ 0.0025 &   2.30$\pm$  0.03 &  0.042$\pm$ 0.010 &  0.96 &  8.39 &  8.45 &  8.30 &  8.38  \\
M101                   & 1323 52797 002 &  0.46 &  0.0196$\pm$ 0.0028 &   5.16$\pm$  0.07 &  0.059$\pm$ 0.004 &  0.94 &  8.47 &  8.35 &  8.59 &  8.40  \\
M101                   & 1324 53088 221 &  0.47 &  0.0177$\pm$ 0.0053 &   5.03$\pm$  0.03 &  0.040$\pm$ 0.012 &  0.92 &  8.43 &  8.37 &  8.54 &  8.42  \\
M101                   & 1323 52797 001 &  0.50 &  0.0073$\pm$ 0.0079 &   2.51$\pm$  0.03 &  0.065$\pm$ 0.012 &  0.87 &  8.59 &  8.40 &  8.65 &  8.43  \\
M101                   & 1324 53088 223 &  0.54 &  0.0566$\pm$ 0.0069 &   7.65$\pm$  0.35 &  0.034$\pm$ 0.015 &  1.15 &  8.23 &  8.23 &  8.25 &  8.25  \\
M101                   & 1323 52797 008 &  0.54 &  0.0135$\pm$ 0.0027 &   4.11$\pm$  0.03 &  0.052$\pm$ 0.004 &  0.90 &  8.47 &  8.36 &  8.59 &  8.41  \\
M101                   & 1324 53088 271 &  0.54 &  0.0138$\pm$ 0.0043 &   2.94$\pm$  0.03 &  0.051$\pm$ 0.004 &  1.00 &  8.39 &  8.38 &  8.34 &  8.34  \\
M101                   & 1324 53088 279 &  0.55 &  0.0352$\pm$ 0.0038 &   5.01$\pm$  0.04 &  0.045$\pm$ 0.005 &  1.13 &  8.22 &  8.25 &  8.20 &  8.23  \\
M101                   & 1324 53088 263 &  0.58 &  0.0587$\pm$ 0.0041 &   6.96$\pm$  0.07 &  0.038$\pm$ 0.005 &  1.21 &  8.17 &  8.19 &  8.17 &  8.19  \\
M101                   & 1325 52762 350 &  0.67 &  0.0386$\pm$ 0.0033 &   6.18$\pm$  0.30 &  0.035$\pm$ 0.004 &  1.09 &  8.25 &  8.25 &  8.27 &  8.27  \\
M101                   & 1323 52797 016 &  0.67 &  0.0343$\pm$ 0.0071 &   4.83$\pm$  0.06 &  0.040$\pm$ 0.008 &  1.14 &  8.20 &  8.26 &  8.16 &  8.23  \\
M101                   & 1324 53088 234 &  0.81 &  0.0686$\pm$ 0.0032 &   6.31$\pm$  0.04 &  0.051$\pm$ 0.003 &  1.33 &   ... &   ... &  8.04 &  ...   \\
M101                   & 1323 52797 066 &  1.05 &  0.0625$\pm$ 0.0120 &   4.91$\pm$  0.06 &  0.054$\pm$ 0.024 &  1.42 &   ... &   ... &  7.89 &  ...   \\
                       &                &       &                     &                   &                   &       &       &       &       &        \\  
NGC428                 &  695 52202 325 &  0.42 &  0.0371$\pm$ 0.0088 &   4.33$\pm$  0.19 &  0.056$\pm$ 0.008 &  1.21 &  8.23 &  8.31 &  8.10 &  8.21  \\
NGC428                 & 1499 53001 453 &  0.43 &  0.0145$\pm$ 0.0080 &   4.25$\pm$  0.02 &  0.039$\pm$ 0.006 &  0.91 &  8.41 &  8.35 &  8.51 &  8.40  \\
NGC428                 & 1499 53001 458 &  0.47 &  0.0230$\pm$ 0.0047 &   4.45$\pm$  0.09 &  0.057$\pm$ 0.004 &  1.03 &  8.36 &  8.29 &  8.39 &  8.30  \\
NGC428                 &  695 52202 324 &  0.51 &  0.0275$\pm$ 0.0118 &   3.96$\pm$  0.02 &  0.045$\pm$ 0.019 &  1.13 &  8.23 &  8.32 &  8.15 &  8.25  \\
NGC428                 &  694 52209 602 &  0.53 &  0.0269$\pm$ 0.0047 &   6.41$\pm$  0.12 &  0.029$\pm$ 0.006 &  0.96 &  8.39 &  8.37 &  8.46 &  8.40  \\
                       &                &       &                     &                   &                   &       &       &       &       &        \\  
NGC450                 &  695 52202 137 &  0.56 &  0.0251$\pm$ 0.0024 &   5.60$\pm$  0.04 &  0.038$\pm$ 0.003 &  0.98 &  8.36 &  8.32 &  8.43 &  8.36  \\
NGC450                 &  398 51789 294 &  0.66 &  0.0170$\pm$ 0.0058 &   3.96$\pm$  0.03 &  0.037$\pm$ 0.005 &  0.97 &  8.32 &  8.32 &  8.36 &  8.34  \\   
                       &                &       &                     &                   &                   &       &       &       &       &        \\  
NGC1110                &  457 51901 304 &  0.10 &  0.0475$\pm$ 0.0124 &   5.65$\pm$  0.05 &  0.040$\pm$ 0.008 &  1.21 &  8.15 &  8.22 &  8.12 &  8.19  \\
NGC1110                &  457 51901 309 &  0.76 &  0.0718$\pm$ 0.0038 &   7.80$\pm$  0.04 &  0.040$\pm$ 0.004 &  1.25 &  8.16 &  8.18 &  8.16 &  8.18  \\
                       &                &       &                     &                   &                   &       &       &       &       &        \\  
NGC2541                &  440 51885 151 &  0.41 &  0.0302$\pm$ 0.0030 &   5.89$\pm$  0.03 &  0.036$\pm$ 0.004 &  1.02 &  8.31 &  8.29 &  8.37 &  8.32  \\
NGC2541                &  440 51912 136 &  0.41 &  0.0277$\pm$ 0.0046 &   5.92$\pm$  0.09 &  0.029$\pm$ 0.049 &  0.99 &  8.33 &  8.32 &  8.38 &  8.35  \\
                       &                &       &                     &                   &                   &       &       &       &       &        \\  
NGC2552                &  440 51885 608 &  0.08 &  0.0149$\pm$ 0.0054 &   3.94$\pm$  0.08 &  0.038$\pm$ 0.008 &  0.94 &  8.37 &  8.33 &  8.44 &  8.37  \\
NGC2552                &  440 51912 627 &  0.08 &  0.0230$\pm$ 0.0085 &   3.57$\pm$  0.03 &  0.048$\pm$ 0.011 &  1.10 &  8.28 &  8.35 &  8.18 &  8.28  \\
                       &                &       &                     &                   &                   &       &       &       &       &        \\  
NGC3023                &  267 51608 384 &  0.41 &  0.0384$\pm$ 0.0041 &   5.49$\pm$  0.03 &  0.035$\pm$ 0.004 &  1.13 &  8.19 &  8.24 &  8.18 &  8.23  \\
NGC3023                &  481 51908 289 &  0.46 &  0.0715$\pm$ 0.0036 &   9.12$\pm$  0.37 &  0.030$\pm$ 0.002 &  1.18 &  8.25 &  8.24 &  8.27 &  8.26  \\
                       &                &       &                     &                   &                   &       &       &       &       &        \\  
NGC3319                & 1361 53047 222 &  0.72 &  0.0373$\pm$ 0.0055 &   6.04$\pm$  0.03 &  0.038$\pm$ 0.005 &  1.08 &  8.26 &  8.25 &  8.29 &  8.27  \\
NGC3319                & 1360 53033 634 &  0.75 &  0.0352$\pm$ 0.0037 &   6.63$\pm$  0.32 &  0.034$\pm$ 0.003 &  1.03 &  8.32 &  8.30 &  8.38 &  8.33  \\
NGC3319                & 1361 53047 221 &  0.81 &  0.0344$\pm$ 0.0031 &   5.80$\pm$  0.17 &  0.050$\pm$ 0.003 &  1.07 &  8.31 &  8.26 &  8.35 &  8.28  \\
                       &                &       &                     &                   &                   &       &       &       &       &        \\  
NGC3432                & 2007 53474 221 &  0.01 &  0.0076$\pm$ 0.0073 &   2.05$\pm$  0.03 &  0.029$\pm$ 0.007 &  0.93 &  8.33 &  8.46 &  8.26 &  8.41  \\
NGC3432                & 2090 53463 550 &  0.14 &  0.0426$\pm$ 0.0033 &   6.27$\pm$  0.23 &  0.033$\pm$ 0.003 &  1.12 &  8.21 &  8.23 &  8.23 &  8.25  \\
NGC3432                & 2090 53463 545 &  0.19 &  0.0138$\pm$ 0.0059 &   2.87$\pm$  0.02 &  0.000$\pm$ 0.000 &  1.00 &   ... &  8.39 &   ... &  8.34  \\
                       &                &       &                     &                   &                   &       &       &       &       &        \\  
NGC3991                & 2095 53474 352 &  0.82 &  0.0158$\pm$ 0.0086 &   3.03$\pm$  0.02 &  0.045$\pm$ 0.005 &  1.03 &  8.33 &  8.38 &  8.25 &  8.32  \\
NGC3991                & 1991 53446 584 &  1.07 &  0.0165$\pm$ 0.0033 &   4.04$\pm$  0.02 &  0.040$\pm$ 0.003 &  0.96 &  8.35 &  8.32 &  8.41 &  8.36  \\
                       &                &       &                     &                   &                   &       &       &       &       &        \\  
NGC3995                & 1991 53446 587 &  0.26 &  0.0270$\pm$ 0.0136 &   3.59$\pm$  0.03 &  0.038$\pm$ 0.012 &  1.16 &  8.18 &  8.38 &  8.05 &  8.25  \\
NGC3995                & 1991 53446 588 &  0.67 &  0.0278$\pm$ 0.0134 &   3.45$\pm$  0.02 &  0.060$\pm$ 0.011 &  1.19 &  8.31 &  8.41 &  8.10 &  8.25  \\
                       &                &       &                     &                   &                   &       &       &       &       &        \\  
NGC4490                & 1452 53112 009 &  0.12 &  0.0166$\pm$ 0.0031 &   4.24$\pm$  0.04 &  0.045$\pm$ 0.003 &  0.94 &  8.39 &  8.33 &  8.47 &  8.37  \\
NGC4490                & 1454 53090 299 &  0.14 &  0.0227$\pm$ 0.0036 &   5.50$\pm$  0.06 &  0.047$\pm$ 0.003 &  0.96 &  8.42 &  8.34 &  8.51 &  8.38  \\
NGC4490                & 1984 53433 366 &  0.16 &  0.0163$\pm$ 0.0037 &   4.03$\pm$  0.02 &  0.039$\pm$ 0.003 &  0.95 &  8.35 &  8.32 &  8.41 &  8.36  \\
NGC4490                & 1452 53112 016 &  0.23 &  0.0248$\pm$ 0.0038 &   4.90$\pm$  0.04 &  0.039$\pm$ 0.003 &  1.02 &  8.30 &  8.28 &  8.34 &  8.31  \\
NGC4490                & 1452 53112 008 &  0.45 &  0.0180$\pm$ 0.0038 &   4.16$\pm$  0.02 &  0.041$\pm$ 0.003 &  0.97 &  8.34 &  8.31 &  8.39 &  8.34  \\
                       &                &       &                     &                   &                   &       &       &       &       &        \\  
NGC5669                & 1709 53533 215 &  0.81 &  0.0350$\pm$ 0.0026 &   7.04$\pm$  0.08 &  0.052$\pm$ 0.004 &  1.01 &  8.41 &  8.33 &  8.49 &  8.36  \\
NGC5669                & 1709 53533 216 &  1.22 &  0.0230$\pm$ 0.0144 &   3.75$\pm$  0.03 &  0.040$\pm$ 0.012 &  1.08 &  8.24 &  8.33 &  8.18 &  8.28  \\
                       &                &       &                     &                   &                   &       &       &       &       &        \\  
IC2828                 & 1223 52781 128 &  0.06 &  0.0239$\pm$ 0.0105 &   4.63$\pm$  0.05 &  0.050$\pm$ 0.006 &  1.03 &  8.33 &  8.28 &  8.36 &  8.30  \\
IC2828                 & 1618 52116 413 &  0.39 &  0.0587$\pm$ 0.0151 &   5.74$\pm$  0.05 &  0.061$\pm$ 0.024 &  1.30 &  8.18 &  8.22 &  8.08 &  8.14  \\
                       &                &       &                     &                   &                   &       &       &       &       &        \\  
UGC05189               & 1305 52757 274 &  0.89 &  0.0280$\pm$ 0.0074 &   4.66$\pm$  0.04 &  0.050$\pm$ 0.006 &  1.07 &  8.29 &  8.27 &  8.28 &  8.27  \\
UGC05189               & 1305 52757 269 &  1.04 &  0.0518$\pm$ 0.0026 &   7.25$\pm$  0.31 &  0.045$\pm$ 0.003 &  1.14 &  8.26 &  8.23 &  8.29 &  8.25  \\
                       &                &       &                     &                   &                   &       &       &       &       &        \\  
UGC05249               &  481 51908 335 &  0.03 &  0.0172$\pm$ 0.0096 &   3.16$\pm$  0.03 &  0.039$\pm$ 0.010 &  1.04 &  8.28 &  8.37 &  8.20 &  8.31  \\
UGC05249               &  480 51989 578 &  0.03 &  0.0109$\pm$ 0.0031 &   3.22$\pm$  0.04 &  0.043$\pm$ 0.005 &  0.91 &  8.42 &  8.36 &  8.48 &  8.39  \\
                       &                &       &                     &                   &                   &       &       &       &       &        \\  
UGC06596               & 1309 52762 012 &  0.09 &  0.0344$\pm$ 0.0092 &   5.04$\pm$  0.03 &  0.047$\pm$ 0.015 &  1.12 &  8.24 &  8.25 &  8.22 &  8.24  \\
UGC06596               & 1311 52765 293 &  0.09 &  0.0262$\pm$ 0.0084 &   5.06$\pm$  0.04 &  0.042$\pm$ 0.010 &  1.03 &  8.31 &  8.28 &  8.35 &  8.31  \\
                       &                &       &                     &                   &                   &       &       &       &       &        \\  
UGC09979               &  315 51663 594 &  0.27 &  0.0314$\pm$ 0.0178 &   3.65$\pm$  0.12 &  0.069$\pm$ 0.022 &  1.21 &  8.33 &  8.40 &  8.11 &  8.23  \\
UGC09979               &  342 51691 352 &  0.27 &  0.0323$\pm$ 0.0121 &   4.01$\pm$  0.03 &  0.054$\pm$ 0.011 &  1.19 &  8.25 &  8.34 &  8.11 &  8.23  \\
                       &                &       &                     &                   &                   &       &       &       &       &        \\  
UGCA154                &  553 51999 596 &  0.05 &  0.0335$\pm$ 0.0089 &   5.51$\pm$  0.03 &  0.038$\pm$ 0.007 &  1.08 &  8.25 &  8.25 &  8.27 &  8.27  \\
UGCA154                &  554 52000 205 &  0.41 &  0.0289$\pm$ 0.0069 &   4.56$\pm$  0.04 &  0.043$\pm$ 0.005 &  1.09 &  8.24 &  8.27 &  8.22 &  8.26  \\
                       &                &       &                     &                   &                   &       &       &       &       &        \\  
UGCA322                &  339 51692 089 &  0.58 &  0.0461$\pm$ 0.0112 &   4.86$\pm$  0.04 &  0.053$\pm$ 0.011 &  1.26 &  8.18 &  8.28 &  8.05 &  8.17  \\
UGCA322                &  339 51692 083 &  0.60 &  0.0341$\pm$ 0.0043 &   6.04$\pm$  0.06 &  0.040$\pm$ 0.003 &  1.05 &  8.30 &  8.27 &  8.35 &  8.30  \\
                       &                &       &                     &                   &                   &       &       &       &       &        \\  
PGC01586               &  390 51816 581 &  0.02 &  0.0370$\pm$ 0.0118 &   4.51$\pm$  0.02 &  0.052$\pm$ 0.010 &  1.20 &  8.22 &  8.29 &  8.12 &  8.21  \\
PGC01586               &  390 51900 596 &  0.02 &  0.0203$\pm$ 0.0057 &   4.34$\pm$  0.05 &  0.053$\pm$ 0.009 &  0.99 &  8.37 &  8.30 &  8.42 &  8.33  \\
                       &                &       &                     &                   &                   &       &       &       &       &        \\  
PGC23706               &  548 51986 517 &  0.02 &  0.0309$\pm$ 0.0064 &   4.50$\pm$  0.02 &  0.056$\pm$ 0.006 &  1.12 &  8.28 &  8.28 &  8.23 &  8.24  \\
PGC23706               &  549 51981 281 &  0.45 &  0.0439$\pm$ 0.0131 &   4.59$\pm$  0.05 &  0.056$\pm$ 0.010 &  1.26 &  8.20 &  8.31 &  8.05 &  8.18  \\
                       &                &       &                     &                   &                   &       &       &       &       &        \\  
PGC51971               &  535 51999 563 &  0.03 &  0.0212$\pm$ 0.0180 &   3.37$\pm$  0.04 &  0.058$\pm$ 0.015 &  1.09 &  8.35 &  8.37 &  8.23 &  8.29  \\
PGC51971               &  585 52027 049 &  0.37 &  0.0300$\pm$ 0.0117 &   4.38$\pm$  0.03 &  0.068$\pm$ 0.013 &  1.12 &  8.33 &  8.29 &  8.27 &  8.25  \\
                       &                &       &                     &                   &                   &       &       &       &       &        \\  
PGC56006               & 1333 52782 613 &  0.01 &  0.0211$\pm$ 0.0081 &   3.81$\pm$  0.02 &  0.051$\pm$ 0.007 &  1.05 &  8.32 &  8.32 &  8.29 &  8.29  \\
PGC56006               & 1168 52731 256 &  0.39 &  0.0242$\pm$ 0.0091 &   3.99$\pm$  0.03 &  0.037$\pm$ 0.007 &  1.08 &  8.22 &  8.31 &  8.19 &  8.27  \\
                       &                &       &                     &                   &                   &       &       &       &       &        \\  
HS1103+4346            & 1363 53053 138 &  0.02 &  0.0556$\pm$ 0.0067 &   6.94$\pm$  0.02 &  0.043$\pm$ 0.006 &  1.18 &  8.20 &  8.20 &  8.21 &  8.21  \\
HS1103+4346            & 1364 53061 295 &  0.51 &  0.0705$\pm$ 0.0053 &   8.14$\pm$  0.07 &  0.043$\pm$ 0.004 &  1.22 &  8.21 &  8.20 &  8.22 &  8.20  \\
                       &                &       &                     &                   &                   &       &       &       &       &        \\  
HS1132+4416            & 1367 53083 292 &  0.01 &  0.0203$\pm$ 0.0102 &   4.77$\pm$  0.03 &  0.043$\pm$ 0.010 &  0.97 &  8.37 &  8.32 &  8.44 &  8.36  \\
HS1132+4416            & 1366 53063 083 &  0.51 &  0.0407$\pm$ 0.0042 &   6.53$\pm$  0.02 &  0.037$\pm$ 0.003 &  1.09 &  8.27 &  8.26 &  8.30 &  8.28  \\
                       &                &       &                     &                   &                   &       &       &       &       &        \\  
UM330                  & 1079 52621 616 &  0.09 &  0.0292$\pm$ 0.0107 &   3.83$\pm$  0.04 &  0.040$\pm$ 0.023 &  1.16 &  8.19 &  8.35 &  8.07 &  8.24  \\
UM330                  & 1078 52643 353 &  0.21 &  0.0289$\pm$ 0.0098 &   5.39$\pm$  0.06 &  0.067$\pm$ 0.014 &  1.04 &  8.38 &  8.28 &  8.44 &  8.30  \\
                       &                &       &                     &                   &                   &       &       &       &       &        \\  
CG1419                 & 1440 53084 103 &  0.03 &  0.0368$\pm$ 0.0063 &   5.99$\pm$  0.03 &  0.041$\pm$ 0.006 &  1.08 &  8.27 &  8.25 &  8.30 &  8.27  \\
CG1419                 & 1996 53436 330 &  0.03 &  0.0458$\pm$ 0.0081 &   6.05$\pm$  0.04 &  0.046$\pm$ 0.008 &  1.16 &  8.21 &  8.22 &  8.21 &  8.22  \\
                       &                &       &                     &                   &                   &       &       &       &       &        \\  
SDSS125446.33+153529.8 & 1771 53498 335 &  0.01 &  0.0513$\pm$ 0.0168 &   6.50$\pm$  0.06 &  0.050$\pm$ 0.021 &  1.18 &  8.22 &  8.21 &  8.22 &  8.21  \\
SDSS125446.33+153529.8 & 1770 53171 612 &  0.01 &  0.0468$\pm$ 0.0169 &   5.93$\pm$  0.06 &  0.045$\pm$ 0.014 &  1.18 &  8.19 &  8.22 &  8.18 &  8.20  \\
                        &                &       &                     &                   &                   &       &       &       &       &        \\  
\enddata
\tablenotetext{a}{Galactocentric distance of the H\,{\sc ii} region normalized to the R$_{25}$ isophotal radius}
\tablenotetext{b}{Dereddened line intensity and its uncertainty on a scale where I$({H\beta})$ = 1}
\tablenotetext{c}{Electron temperature within the [O III] 
zone in units of 10$^4$K}
\tablenotetext{d}{t$_2$ abundances 
calculated with Eq. (\ref{equation:ttp}) and given as 
12 + log (O/H)}
\tablenotetext{e}{t$_2$ abundances calculated with Eq. (\ref{equation:tt1})
and given as 12 + log (O/H)} 
\end{deluxetable}


\clearpage


\begin{deluxetable}{lrccc}
\tablewidth{0pc}
\tablecaption{Comparison of (O/H)$_{7325}$ abundances with those of other 
authors\label{table:compar}}
\tabletypesize{\footnotesize}
\tablehead{
galaxy name                                              & 
spectrum number                                          & 
(O/H)$_{7325}$\tablenotemark{a}                          & 
(O/H)$_{7325}$\tablenotemark{a}                          & 
(O/H)$_{7325}$\tablenotemark{a}                          \\
                                                         & 
                                                         & 
this study                                               & 
Izotov et al. (2006)                                     & 
Kniazev et al (2004)                                     }                              
\startdata 
  M101         &     1325 52762 348   &     8.24   &     8.26   &           \\
  M101         &     1325 52762 352   &     8.58   &     8.48   &           \\
  M101         &     1323 52797 002   &     8.59   &     8.51   &           \\
  M101         &     1323 52797 008   &     8.59   &     8.53   &           \\
  M101         &     1325 52762 350   &     8.27   &     8.22   &           \\
  M101         &     1323 52797 016   &     8.16   &     8.07   &           \\
  NGC450       &      398 51789 294   &     8.36   &     8.40   &     8.18  \\
  NGC1110      &      457 51901 304   &     8.12   &     8.13   &     8.06  \\
  NGC1110      &      457 51901 309   &     8.16   &     8.03   &     8.03  \\
  NGC2541      &      440 51885 151   &     8.37   &     8.30   &     8.23  \\
  NGC2552      &      440 51885 608   &     8.44   &     8.39   &     8.28  \\
  NGC3023      &      267 51608 384   &     8.18   &     7.99   &     8.10  \\
  NGC3023      &      481 51908 289   &     8.27   &     8.12   &     8.08  \\
  IC2828       &     1223 52781 128   &     8.36   &     8.22   &           \\
  UGC05189     &     1305 52757 269   &     8.29   &     8.24   &           \\
  UGC05249     &      481 51908 335   &     8.20   &     8.23   &           \\
  UGCA154      &      553 51999 596   &     8.27   &     8.22   &           \\
  UGCA154      &      554 52000 205   &     8.22   &            &     8.15  \\
  UGCA322      &      339 51692 089   &     8.05   &            &     7.96  \\
  UGCA322      &      339 51692 083   &     8.35   &     8.15   &     8.23  \\
  PGC001586    &      390 51900 596   &     8.42   &     8.38   &     8.21  \\
  PGC023706    &      548 51986 517   &     8.23   &     8.23   &     8.05  \\
  PGC023706    &      549 51981 281   &     8.05   &     8.16   &     8.13  \\
  PGC051971    &      535 51999 563   &     8.23   &     8.30   &           \\
  PGC051971    &      585 52027 049   &     8.27   &     8.10   &     8.18  \\
  PGC056006    &     1333 52782 613   &     8.29   &     8.26   &           \\
\enddata
\tablenotetext{a}{Oxygen abundances are given as 12 + log (O/H); they are
derived with the one-dimensional t$_2$ -- t$_3$ relation}
\end{deluxetable}


\clearpage


\begin{deluxetable}{lccc}
\tablewidth{0pc}
\tablecaption{Absolute blue luminosity and central metallicity 
of galaxies in nearby sample\label{table:lummet}}
\tabletypesize{\footnotesize}
\tablehead{
galaxy                                                   & 
M$_{B}$                                                  & 
central (O/H)\tablenotemark{a}                           & 
central (O/H)\tablenotemark{a}                          \\
name                                                     & 
                                                         & 
t$_2$ from Eq.(16)                                       & 
t$_2$ from Eq.(15)                                       }                              
\startdata 
NGC1110         &  -18.17  &  8.19  &  8.15   \\
NGC2552         &  -17.97  &  8.33  &  8.33   \\
NGC4490         &  -21.55  &  8.36  &  8.43   \\
IC2828          &  -16.49  &  8.26  &  8.24   \\
UGC05249        &  -18.95  &  8.36  &  8.36   \\
UGC06569        &  -16.28  &  8.27  &  8.28   \\
UGCA154         &  -18.48  &  8.25  &  8.26   \\
PGC01586        &  -17.68  &  8.30  &  8.28   \\
PGC23706        &  -17.25  &  8.27  &  8.18   \\
PGC51971        &  -17.02  &  8.34  &  8.26   \\
PGC56006        &  -18.28  &  8.29  &  8.26   \\
\enddata
\tablenotetext{a}{Oxygen abundances are given as 12 + log (O/H)}
\end{deluxetable}


\newpage


\newpage


\newpage


\newpage


\newpage


\newpage


\newpage


\newpage


\newpage


\newpage


\end{document}